\documentclass[useAMS,usenatbib]{mn2e}
\usepackage{graphicx}
\usepackage{color}
\usepackage{times}
\usepackage{amsmath,amssymb}
\usepackage[total={17.8cm,24.0cm},centering]{geometry}
\title[Slow and Fast Rotators in the Fornax Cluster]
{Distribution of Slow and Fast Rotators in the Fornax Cluster}
\author[N. Scott et al.]{Nicholas Scott,$^{1,2,3}$\thanks{E-mail: nscott@physics.usyd.edu.au}, Roger L Davies$^4$, Ryan C W Houghton$^4$, Michele Cappellari$^4$,\newauthor Alister W Graham$^3$ and Kevin A Pimbblet$^{4,5,6}$\\
$^1$Sydney Institute for Astronomy (SIfA), School of Physics, The University of Sydney, NSW 2006, Australia\\
$^2$ARC Centre of Excellence for All-sky Astrophysics (CAASTRO)\\
$^3$Centre for Astrophysics \& Supercomputing, Swinburne University of Technology, PO Box 218, Hawthorn, VIC 3122, Australia\\
$^4$Sub-department of Astrophysics, Department of Physics, University of Oxford, Denys Wilkinson Building, Keble Road, Oxford, OX1 3RH, UK\\
$^5$School of Physics, Monash University, Clayton, VIC 3800, Australia\\
$^6$Department of Physics and Mathematics, University of Hull, Cottingham Road, Hull, HU6 7RX, UK}
\date{6$^th$ December, 2013}
\pagerange{\pageref{firstpage}--\pageref{lastpage}} \pubyear{2013}

\def\LaTeX{L\kern-.36em\raise.3ex\hbox{a}\kern-.15em
    T\kern-.1667em\lower.7ex\hbox{E}\kern-.125emX}

\begin{document}

\label{firstpage}

\maketitle

\begin{abstract}
We present integral field spectroscopy of 10 early-type galaxies in the nearby, low-mass, Fornax cluster, from which we derive spatially resolved stellar kinematics. Based on the morphologies of their stellar velocity maps we classify 2/10 galaxies as slow rotators, with the remaining 8 galaxies fast rotators.

Supplementing our integral field observations with morphological and kinematic data from the literature, we analyse the `kinematic' type of all 30 galaxies in the Fornax cluster brighter than M$_K = -21.5$ mag (M$_* \sim 6 \times 10^9$ M$_\odot$). Our sample's slow rotator fraction within one virial radius is $7^{+4}_{-6}$ per cent. $13^{+8}_{-6}$ per cent of the early-type galaxies are slow rotators, consistent with the observed fraction in other galaxy aggregates. The fraction of slow rotators in Fornax varies with cluster-centric radius, rising to 16$^{+11}_{-8}$ per cent of all kinematic types within the central 0.2 virial radii, from 0 per cent in the cluster outskirts. 

We find that, even in mass-matched samples of slow and fast rotators, slow rotators are found preferentially at higher projected environmental density than fast rotators. This demonstrates that dynamical friction alone cannot be responsible for the differing distributions of slow and fast rotators. For dynamical friction to play a significant role, slow rotators must reside in higher mass sub-halos than fast rotators and/or form in the centres of groups before being accreted on to the cluster.
\end{abstract}


\section{Introduction}
\label{sec:intro}
The population of galaxies that reside in clusters shows significant differences to that in the field and other lower-density environments. Cluster galaxies are more likely to be red \citep{Butcher:1978a} and have a higher fraction of early-type morphologies \citep{Oemler:1974,Dressler:1980} than similar populations in low density environments. Cluster galaxies also typically have lower specific star formation rates than group and field galaxies. At higher redshifts the difference between cluster and field galaxies is less pronounced, with the red fraction of cluster galaxies increasing significantly since z\ $=0.5$ \citep{Butcher:1978b}. A number of authors \citep{Stanford:1998,vanDokkum:2000,Smith:2005,Postman:2005,Cooper:2006,Capak:2007,Poggianti:2008} have more recently confirmed this picture, showing that in the highest density regions the early-type fraction is already high at z\ $\sim 1$ (though increases still further up to the present day), but in low and intermediate density regions significant morphological evolution is only seen between z\ $=0.5$ and today.

Despite these long-recognised differences the precise effects of environment on galaxy evolution are poorly understood. Much of this uncertainty is due to the difficulty of disentangling the effects of galaxy mass from environment. Many of the differences between cluster and non-cluster galaxy populations are similar to differences between high and low-mass galaxy populations. It has been suggested that the observed effects of environment are simply due to a differing mass function between low- and high-density environments \citep{Treu:2005}, although recent studies involving large samples do find a statistically significant variation in e.g. specific star formation rate beyond that which would be expected from a differing mass function alone \citep{Bamford:2009,Peng:2010}.

While large samples are effective in quantifying the difference between cluster and non-cluster galaxy populations they are less effective at revealing the physical mechanisms responsible for those differences. Many such mechanisms have been proposed, including: ram pressure stripping \citep{Gunn:1972}, strangulation \citep{Larson:1980}, galaxy-galaxy mergers, pre-processing in infalling groups, harassment \citep{Moore:1996} and tidal interactions \citep[see][for an overview]{Boselli:2006}. Individual examples of many of these processes in action have been observed \citep[e.g.][who identified galaxies undergoing ram-pressure stripping in the Virgo cluster and Shapley supercluster respectively]{Chung:2007,Chung:2009,Abramson:2011,Merluzzi:2013}. What is missing is an understanding of how frequently these processes occur and which are significant in driving the observed differences between cluster and non-cluster galaxies.

A fruitful compromise between detailed studies of individual objects and large surveys that lack high-quality data are integral field spectrograph (IFS) surveys of nearby clusters. By combining the new insights derived from two-dimensional spectroscopic information with the statistical power of large samples we can shed new light on the detailed role of environment in shaping the way galaxies evolve. \citet{Emsellem:2007} and \citet{Cappellari:2007} proposed a kinematic classification for early-type galaxies (E and S0) based on the morphology of their velocity maps, dividing early-type galaxies into two kinematic classes. These classes are slow rotators (SRs), systems with low specific angular momentum, and fast rotators (FRs), systems with significant, ordered, disk-like rotation. This classification is less affected by the inclination affects that trouble visual morphology classification schemes and seems likely related to the formation history of each galaxy. It is also complementary to structural morphology classifications obtained from quantitative bulge-disc decompositions.

\citet{Cappellari:2011b} applied this classification scheme \citep[as updated by][]{Krajnovic:2011,Emsellem:2011} to revisit the morphology-density relation, presenting the kinematic morphology-density relation for the ATLAS$^\mathrm{3D}$ survey --- a large sample of early-type galaxies brighter than M$_K = -21.5$ mag covering a range of environments from the Virgo cluster to galaxy groups and the field \citep{Cappellari:2011a}. They found that spirals are transformed into FRs at a rate that increases linearly (with log environment) across all environments, whereas the ratio of SRs to FRs seems nearly insensitive to environment. Only in the Virgo cluster was a clear difference observed. There, all SRs are found within the dense core of the Virgo cluster. The Coma cluster was studied by \citet{Scott:2012} and \citet{Houghton:2013}, and the Abell~1689 cluster by \citet{DEugenio:2013} using a range of integral-field spectrographs. Both clusters are significantly more massive than the Virgo cluster.  They again found that, within a cluster, the SR fraction increases significantly in the densest regions, however the local environmental density that this increase occurs at varies widely from cluster to cluster. \citet{DEugenio:2013} and \citet{Houghton:2013} argued that SRs make up a constant proportion, $\sim 15$ per cent, of early-type galaxies brighter than M$_{K} = -21.5$ mag, independent of the environment explored. 

Past studies have focused predominantly on massive clusters or the field, with the low-mass cluster and group environment largely unexplored. In this paper we present an integral field study of early-type galaxies in the Fornax cluster. Fornax is a relatively low-mass cluster \citep[M$_* \sim 7 \times 10^{13}$ M$_\odot$][approximately 1/10$^{th}$ the mass of the Virgo cluster]{Drinkwater:2001}, lying close to the boundary between cluster and group environments. Adding the Fornax cluster to the previously mentioned IFS studies of cluster early-type galaxies extends coverage of the broad range of cluster environments found in the nearby Universe.

Several works have already examined the spatially-resolved kinematics of bright galaxies in the Fornax cluster, though these have been limited to long-slit spectroscopy observations. Of these, \citet{DOnofrio:1995} and \citet{Graham:1998} present the largest number of galaxies (15 and 12 respectively). \citet{DOnofrio:1995} discovered that 6 of their 9 galaxies classified as elliptical harbour a disc-like component. Building on this, \citet{Graham:1998} reported that only 3 of their 12 brightest (M$_B \le 14.7$ mag, $M_K \le 11.8$ mag) elliptically-classified galaxies in the Fornax cluster are actually pressure supported systems. Long-slit studies can accurately identify FRs and SRs when the galaxy is significantly flattened, or when either the rotation or dispersion is highly dominant. In intermediate cases the spatial information contained in IFS data is critical in accurately classifying the kinematic morphology of a galaxy. The classic case of this is identifying face-on discs, which traditional classification schemes often but mistakenly class as ellipticals \citep{Emsellem:2011}. The key advance of this study over past long-slit work is in using IFS data to accurately separate disc-like systems from true SRs, and then applying this robust kinematic classification to the morphology--density relation in the low-mass cluster environment of Fornax.

In Section \ref{sec:sample} we describe our sample of Fornax galaxies. In Sections \ref{sec:spec} and \ref{sec:phot} we describe the integral field observations, reduction and kinematic analysis, as well as supplementary information derived from imaging. In Section \ref{sec:results} we present the kinematic morphology-density relation for the Fornax cluster. In Section \ref{sec:disc} we place our Fornax cluster results into context with other IFS surveys of nearby clusters and discuss the implications for the impact of cluster environments on the evolution of early-type galaxies.

\section{Sample}
\label{sec:sample}

\begin{table*}
\caption{Early-type galaxies in the Fornax Cluster}
\label{tab:table1}
\begin{center}
\begin{tabular}{l c c c c c c c c}
\hline
Object & RA                & DEC            & m$_K$ & $\mu_B$ & $\epsilon_\mathrm{e}$ & $\lambda_{R_e}$ &Kinematic & log $\Sigma_3$ \\
             & (hh:mm:ss) & (dd:mm:ss) & (mag)   & (mag arcsec$^{-2}$)    & & &Type & (Mpc$^{-2}$) \\
 (1) & (2) & (3) & (4) & (5) & (6) & (7) & (8) & (9)\\             
\hline
\hline
IC~1963         & 03:35:31.0 & -34:26:50 & 9.15 & 19.7 & 0.69 & -- & (F) & 1.11\\
IC~2006         & 03:54:28.4 & -35:58:02 & 8.48 & 21.5 & 0.13 & 0.37 & F & 0.21\\
NGC~1316    & 03:22:41.7 & -37:12:30 & 7.74 & 20.9 & 0.31 & -- & (F) & 1.11\\
NGC~1336    & 03:26:32.2 & -35:42:49 & 9.81 & 22.5 & 0.27 & 0.32 & F & 0.56\\
NGC~1339    & 03:28:06.6 & -32:17:10 & 8.69 & 20.7 & 0.27 & 0.49 & F & 0.46\\
NGC~1351    & 03:30:35.0 & -34:51:11 & 8.79 & 21.0 & 0.35 & -- & (F) & 0.82\\
NGC~1374    & 03:35:16.6 & -35:13:35 & 8.16 & 20.9 & 0.10 & 0.24 & F & 2.04\\
NGC~1375    & 03:35:16.8 & -35:15:57 & 9.61 & 21.3 & 0.56 & -- & (F) & 2.08\\
NGC~1379    & 03:36:03.9 & -35:26:28 & 8.24 & 21.8 & 0.05 & 0.25 & F & 2.14\\
NGC~1380    & 03:36:27.6 & -34:58:35 & 6.87 & 20.8 & 0.51 & -- & (F) & 1.89\\
NGC~1380A & 03:36:47.5 & -34:44:23 & 9.57 & 21.8 & 0.72 & -- & (F) & 1.57\\
NGC~1381    & 03:36:31.7 & -35:17:43 & 8.42 & 19.8 & 0.62 & -- & (F) & 2.18\\
NGC~1382    & 03:37:08.9 & -35:11:42 & 10.0 & 20.9 & 0.05 & 0.09 & F & 1.90\\
NGC~1387    & 03:36:57.0 & -35:30:24 & 7.43 & 20.4 & 0.05 & -- & (F) & 2.12\\
NGC~1389    & 03:37:11.8 & -35:44:46 & 8.63 & 20.3 & 0.44 & -- & (F) & 1.75\\
NGC~1399    & 03:38:29 & -35:27:03 & 6.31 & 20.6 & 0.09 & 0.08 & S & 1.75\\
NGC~1404    & 03:38:51.9 & -35:35:40 & 6.82 & 19.9 & 0.12 & 0.20 & F & 1.69\\
NGC~1419    & 03:40:42.1 & -37:30:39 & 9.89 & 20.4 & 0.01 & 0.19 & F & 0.40\\
NGC~1427    & 03:42:19.4 & -35:23:34 & 8.14 & 21.2 & 0.31 & 0.16 & S & 1.16\\ 
NGC~1460    & 03:46:13.7 & -36:41:47 & 9.96 & -- & 0.15 & -- & (F) & 0.53\\
\hline
\end{tabular}
\end{center}
Notes: Columns (1), (2) and (3): From the NASA Extagalactic Database (NED). Column (4): From the 2MASS Extended Source Catalogue \citep{Jarrett:2000}. Column(5): From the HyperLeda database \citep{Paturel:2003}. Column (8): Brackets indicate classifications based on literature kinematic data or the apparent ellipticity. See Section \ref{sec:without} for details.
\end{table*}

Our initial sample consisted of a complete, volume-limited sample of early-type galaxies in the Fornax cluster with absolute magnitude brighter than M$_{Ks} = -21.5$ mag. This absolute magnitude limit is the same as that used in the ATLAS$^\mathrm{3D}$ survey, and was chosen to facilitate comparison with the results of that survey. The limit corresponds to a stellar mass selection M$_* \ga 6\times10^9$ M$_\odot$, assuming a typical stellar mass-to-light ratio for early-type galaxies of $\sim 0.75$ in the $K_s$ band.

Our sample was selected from the Two Micron All Sky Survey (2MASS) Extended Source Catalogue \citep{Jarrett:2000}. We began by selecting all galaxies within four degrees of RA = 03:38:29.0 DEC = -35:27:03, the position of NGC~1399, generally accepted to lie at the centre of the cluster. Four degrees corresponds to a physical radius of 1.4 Mpc at the distance of Fornax, \citep[20.0 Mpc,][]{Blakeslee:2009}, which is the Virial radius of the cluster as estimated by \citet{Drinkwater:2001}. We then eliminated all objects fainter than m$_{Ks}  = 10.1$ mag, which corresponds to an absolute magnitude M$_{Ks} = -21.5$ mag at the distance of Fornax. Late-type galaxies containing spiral arms were then eliminated from the sample by visual inspection of 2MASS K$_\mathrm{s}$-band and Digital Sky Survey B-band images by NS. This yielded a complete cluster sample of 20 early-type galaxies, which are listed in Table \ref{tab:table1}. For clarity, we define the 30 cluster members brighter than M$_{Ks} = -21.5$ mag as the parent sample, and the 20 members of the parent sample with an early-type visual morphology as the early-type sample. In the following section we describe the IFS observations of 10 targets from the early-type sample.

\section{Spectroscopic data}
\label{sec:spec}

We obtained integral field spectroscopy of our sample using the Wide Field Spectrograph \citep[WiFeS][]{Dopita:2010} on the Australian National University 2.3m telescope. WiFeS is an integral field spectrograph based on the image slicing concept, with 25 slices of length 76 pixels. The instrument has a field-of-view of $25 \times 38$ arcseconds, with 1 arcsecond square pixels (after binning two pixels along each slice). This corresponds to $2.4 \times 3.7$ kpc at the distance of Fornax. We used the B3000 grating that provides wavelength coverage from $\sim$ 3600 -- 5700 \AA, with a spectral resolution of full-width at half-maximum of $\sim 2.2$ \AA\ ($\sigma_{inst} \sim 62$ km s$^{-1}$). The instrument provides simultaneous observations at the red end of the optical wavelength range, however we do not use that data in this study and do not describe the red spectra here.

\subsection{WiFeS observations}

\begin{table}
\caption{Summary of WiFeS observations}
\label{tab:table2}
\begin{center}
\begin{tabular}{l l c c}
\hline
Object & Run Date & Exp. Time (s) & Exposures \\
\hline
\hline
IC~2006      & 17$^\mathrm{th}$ Oct 2012 & 7200 & 3 \\
NGC~1336 & 17$^\mathrm{th}$ Oct 2012 & 7200 & 3 \\
NGC~1336 & 16$^\mathrm{th}$ Nov 2012 & 7200 & 3 \\
NGC~1339 & 16$^\mathrm{th}$ Nov 2012 & 3600 & 2 \\
NGC~1374 & 17$^\mathrm{th}$ Oct 2012 & 6600 & 3 \\
NGC~1379 & 17$^\mathrm{th}$ Oct 2012 & 7200 & 4 \\
NGC~1382 & 16$^\mathrm{th}$ Nov 2012 & 7200 & 3 \\
NGC~1399 & 17$^\mathrm{th}$ Oct 2012 & 7200 & 4$^\dagger$ \\
NGC~1404 & 16$^\mathrm{th}$ Nov 2012 & 7200 & 3 \\
NGC~1419 & 17$^\mathrm{th}$ Oct 2012 & 7200 & 3 \\
NGC~1427 & 17$^\mathrm{th}$ Oct 2012 & 7200 & 3 \\
NGC~7710 & 17$^\mathrm{th}$ Oct 2012 & 7200 & 4 \\
\hline
\end{tabular}
\end{center}
$^\dagger$ For NGC~1399 the 4 exposures were evenly divided between two pointings either side of the galaxy centre.
\end{table}

Our WiFeS observations were carried out over two separate observing runs, the 17$^\mathrm{th}$--21$^\mathrm{st}$ of October, 2012 and the 16$^\mathrm{th}$--18$^\mathrm{th}$ of November, 2012. Due to time constraints and time lost due to weather we were only able to observe 10 of our targets. We prioritised targets with ellipticities rounder than $\epsilon = 0.4$ in the HyperLeda database\footnote{http://leda.univ-lyon1.fr} \citep{Paturel:2003}, as flatter objects were shown to be almost certainly FRs \citep{Emsellem:2011} or in some cases counter-rotating discs \citep{Krajnovic:2011}. We additionally observed NGC~7710, a galaxy also observed as part of the ATLAS$^\mathrm{3D}$ survey, to verify the quality of our stellar kinematics. For the majority of targets we obtained total integration times of 2 hours, consisting of 3 or 4 separate exposures dithered by roughly one pixel. For NGC~1399, the largest galaxy in our sample, we obtained two 1 hour pointings offset by 20 arcseconds either side of the centre of the galaxy. These pointings consisted of two 30 minute exposures, again dithered by roughly one pixel. The seeing, as estimated from observations of standard stars, was typically $\sim 1.5$ arcseonds. The start date of the observing run each object was observed on, the total exposure time and the number of pointings are given in Table \ref{tab:table2}. In addition to the science object exposures, we took separate sky exposures of duration 15 or 20 minutes adjacent to each science object exposure. We also obtained arc frames at the beginning and end of each object-sky sequence, and the standard calibration data of bias, lamp flat, twilight flat and wire frames at the beginning of each night.

\begin{figure*}
\includegraphics[width=6.75in,clip,trim = 10 10 20 20]{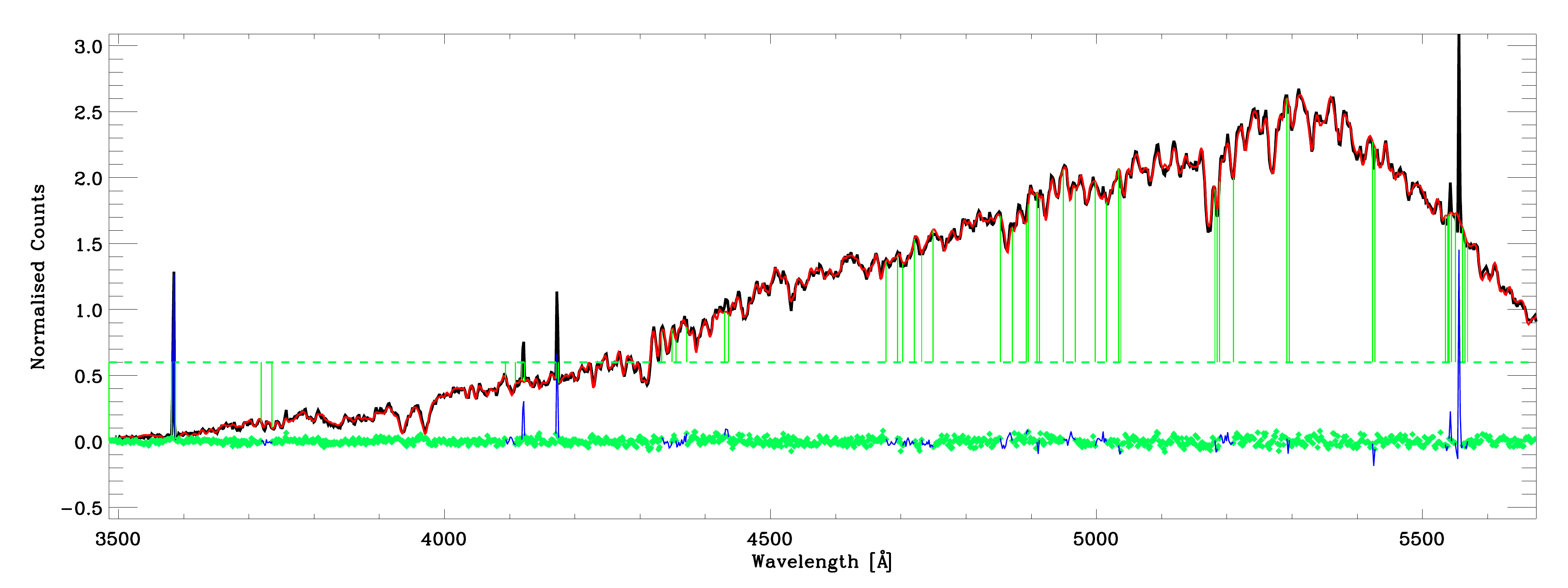}
\caption{Example of the pPXF fit to the central spectrum for NGC~1374. The observed spectrum, formed by binning all spaxels in the central 5'', is shown in black, with the best-fitting pPXF model shown in red. The residuals of the best-fitting model are shown as the green dots. The vertical green lines indicate regions that were excluded from the fit due to emission features or large residuals, with the blue lines showing the residuals between the observed spectrum and model in these regions. The spectrum shown is not flux calibrated.}
\label{fig:figure2}
\end{figure*}

\subsection{WiFeS data reduction}

Our data were reduced using the Python-based data reduction pipeline for WiFeS, PyWiFeS\footnote{http://www.mso.anu.edu.au/pywifes/} \citep{Childress:2013}. This involves the standard steps of bias subtraction, slitlet extraction, cosmic ray removal, flat-fielding, wavelength calibration, differential atmospheric refraction and data cube construction. From each spatial pixel in each science exposure we subtracted a single high S/N sky spectrum formed by spatially collapsing a sky data cube observed adjacent in time to the science exposure. This resulted in a sky subtraction accurate to 1 per cent across the full field-of-view and wavelength range of the instrument, except for around the prominent sky line at 5577 \AA. This region was masked in all subsequent steps of the reduction and analysis due to the significant sky line residual.

Individual science exposures were combined into a single science data cube using a purpose-built Python routine. The offsets between the individual frames were determined by first collapsing each cube along the wavelength direction to form an image. A centroid for each image was then determined using a Python routine adapted from the IDL routine {\it find\_galaxy.pro}\footnote{Available as part of the mge\_fit\_sectors package \citep{Cappellari:2002} from http://purl.org/cappellari/idl}. This routine determines the centroid of an image using the second moments of the luminosity distribution to sub-pixel accuracy. Finally, the individual data cubes were combined using the sub-pixel shifts determined in the previous step.

\subsection{Stellar Kinematic Extraction}
\label{sec:maps}

The extraction of robust stellar velocities and velocity dispersions requires high S/N data. For spatial pixels in the centre of our targets the S/N per pixel is typically greater than 60, more than sufficient for our purposes. However, in the outer parts of the field-of-view the S/N per pixel falls to $\sim$ 1--3, insufficient to extract reliable stellar kinematics. To increase the S/N in the outer parts of our galaxies we made use of the Voronoi binning technique, as implemented by \citet{Cappellari:2003}. The Voronoi technique accretes spatial pixels onto a bin until the target S/N is reached, ensuring all spatial pixels are included and all bins are contiguous, though the outermost bins may not reach the target S/N if insufficient spatial pixels are available. We adopt a target S/N of 40 per bin, as a compromise between maximising the effective spatial resolution and coverage of each object while still ensuring sufficiently high S/N to extract reliable stellar kinematics.

Stellar kinematics were determined using the penalised Pixel Fitting (pPXF) routine of \citet{Cappellari:2004}. Given a set of template spectra, this routine simultaneously determines the optimal combination of the input templates and the best-fitting line of sight velocity distribution (LOSVD) for an input spectrum. For our stellar template spectra we elected to use the Indo-US Coud\'e Feed Spectral Library of 1273 stars from \citet{Valdes:2004}, as this provided the best match to the wavelength range and spectral resolution of our observations. 

For each galaxy we first extracted a high S/N ($\gtrsim 100$) spectrum by binning all spatial pixels within 5 arcseconds of the object's centre. From this high S/N spectrum we determined the best-fitting template spectrum (typically consisting of a linear combination of $\sim 20$ of the stellar templates), systemic velocity and central velocity dispersion. We include an additive polynomial of 10$^\mathrm{th}$ degree in each fit, to account for instrumental effects on the overall shape of the spectrum and due to the continuum shape not being flux calibrated. We also fit for the higher-order Gauss-Hermite moments, h$_\mathrm{3}$ and h$_\mathrm{4}$, though we were not able to measure these reliably in all cases. The BIAS parameter was optimised using simulated data of the appropriate dispersion and S/N following \citet{Cappellari:2011a}. The spectral region around a number of prominent emission lines was masked in each fit, even in cases where no obvious emission was present. We used pPXF in an iterative way with the /CLEAN keyword, to reject any remaining bad pixels or unmasked emission. An example of this central spectrum (in black) with the best-fitting pPXF model (in red) is shown in Figure \ref{fig:figure2}.

\begin{figure*}
\includegraphics[width=6in,clip,trim= 100 25 105 35]{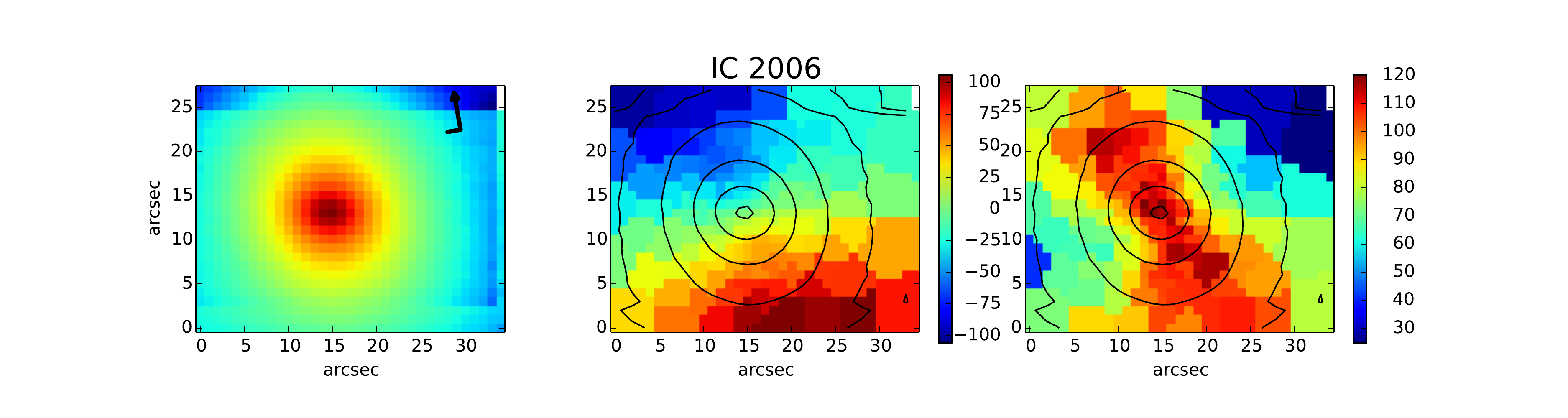}
\includegraphics[width=6in,clip,trim= 100 25 105 35]{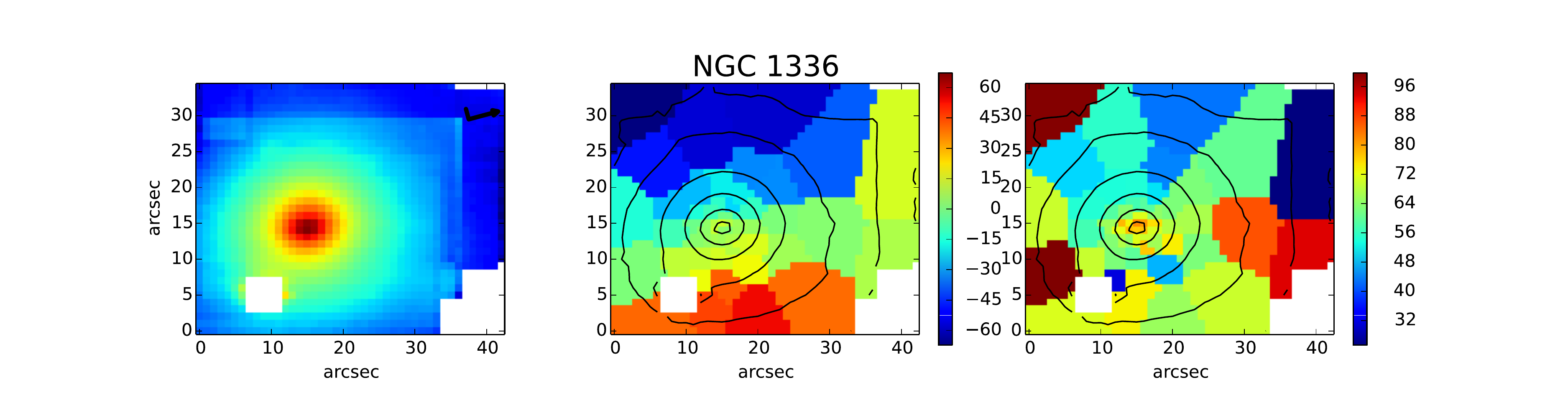}
\includegraphics[width=6in,clip,trim= 100 33 105 45]{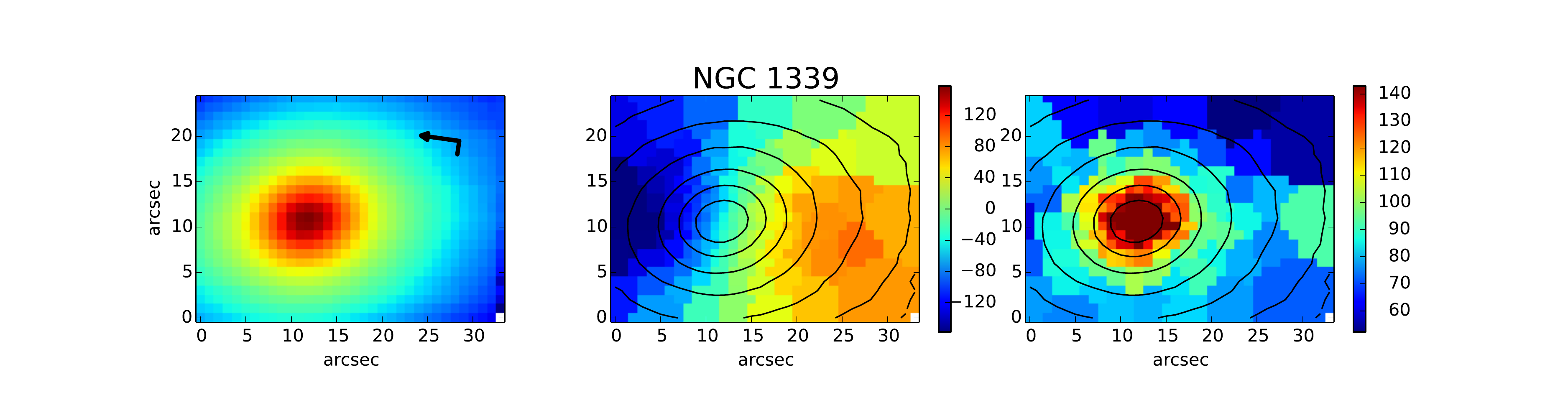}
\includegraphics[width=6in,clip,trim= 100 29 105 45]{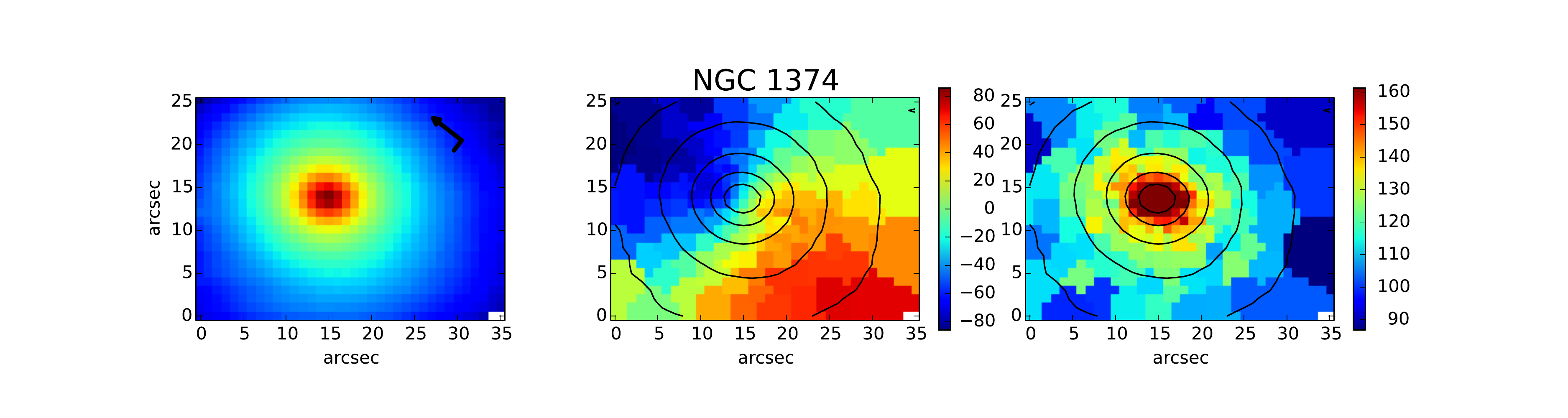}
\includegraphics[width=6in,clip,trim= 100 30 105 45]{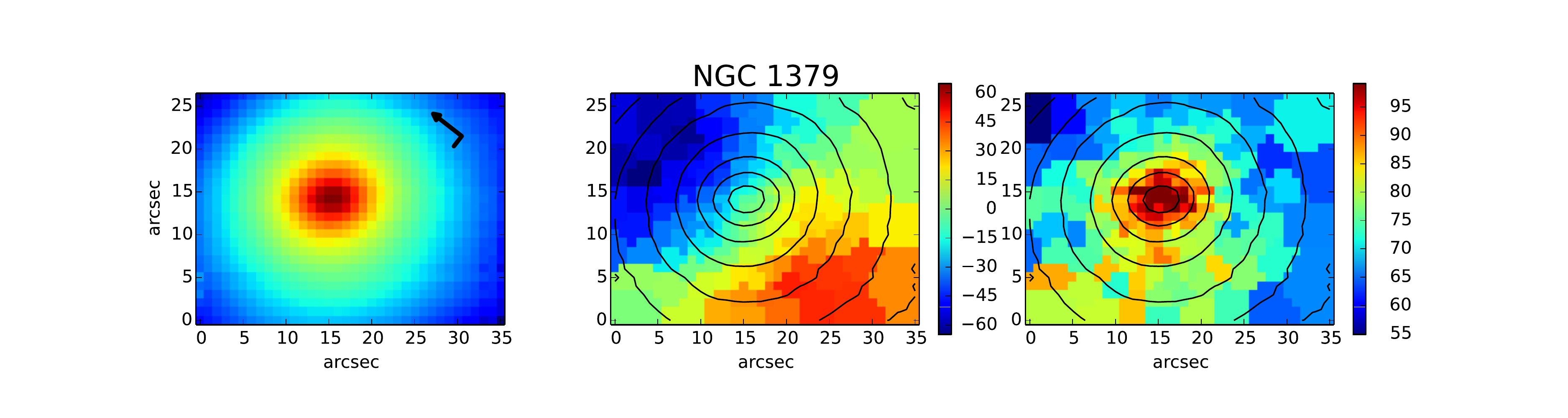}
\includegraphics[width=6in,clip,trim= 100 30 100 45]{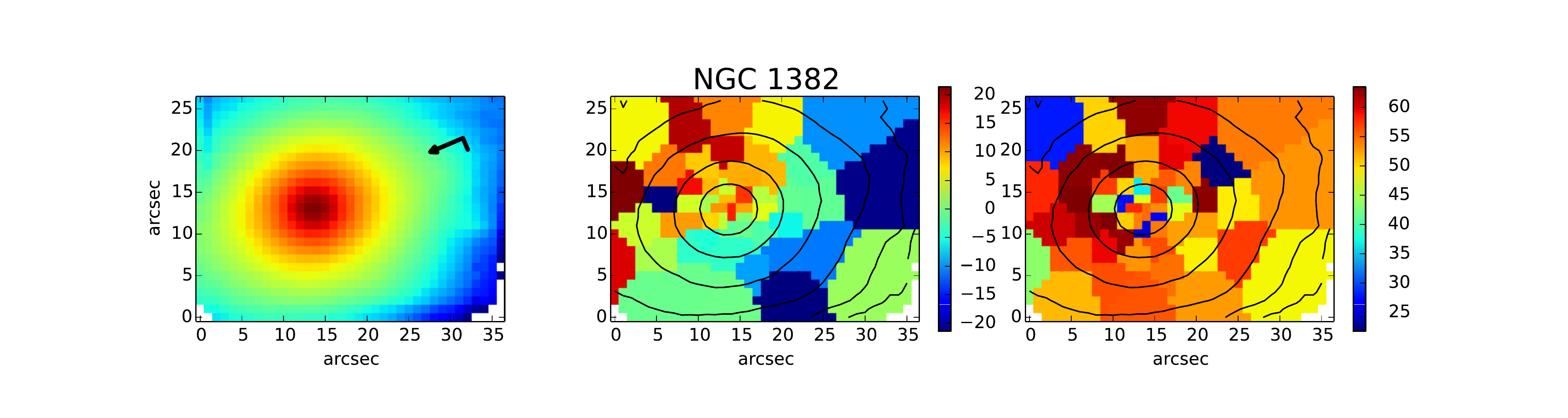}
\caption{Flux, mean stellar velocity and mean stellar velocity dispersion maps for the 10 galaxies for which we obtained WiFeS integral field observations. Foreground stars have been masked for NGC~1399 and NGC~1336. North is indicated by the long arrow, with east indicated by the shorter dash. The velocity scale for the velocity and dispersion maps is in km s$^{-1}$.}
\label{fig:figurea1}
\end{figure*}

\begin{figure*}
\includegraphics[width=6in,clip,trim= 120 0 90 10]{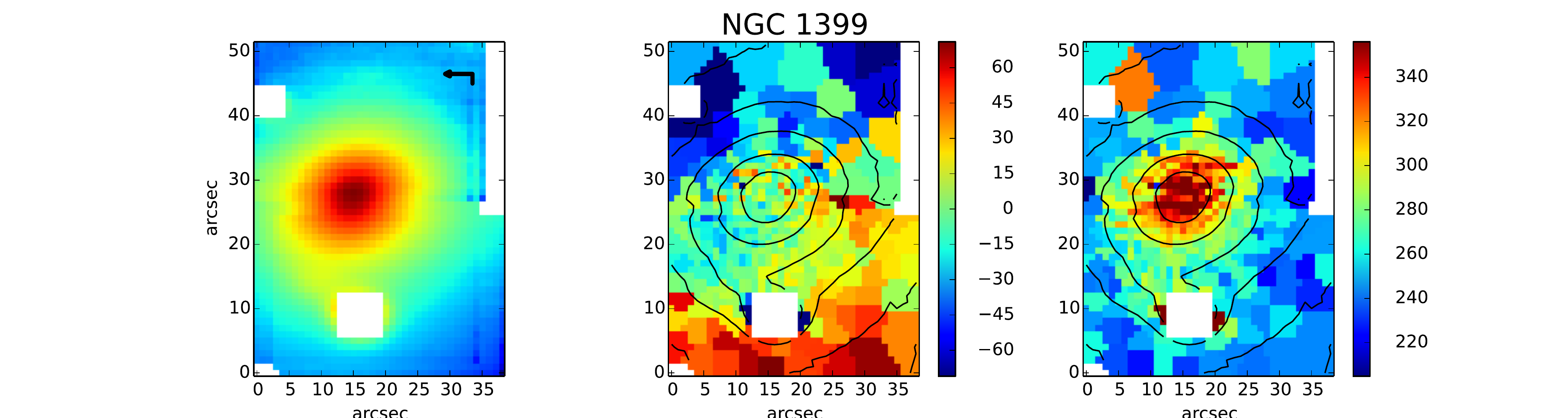}
\includegraphics[width=6in,clip,trim= 100 30 100 40]{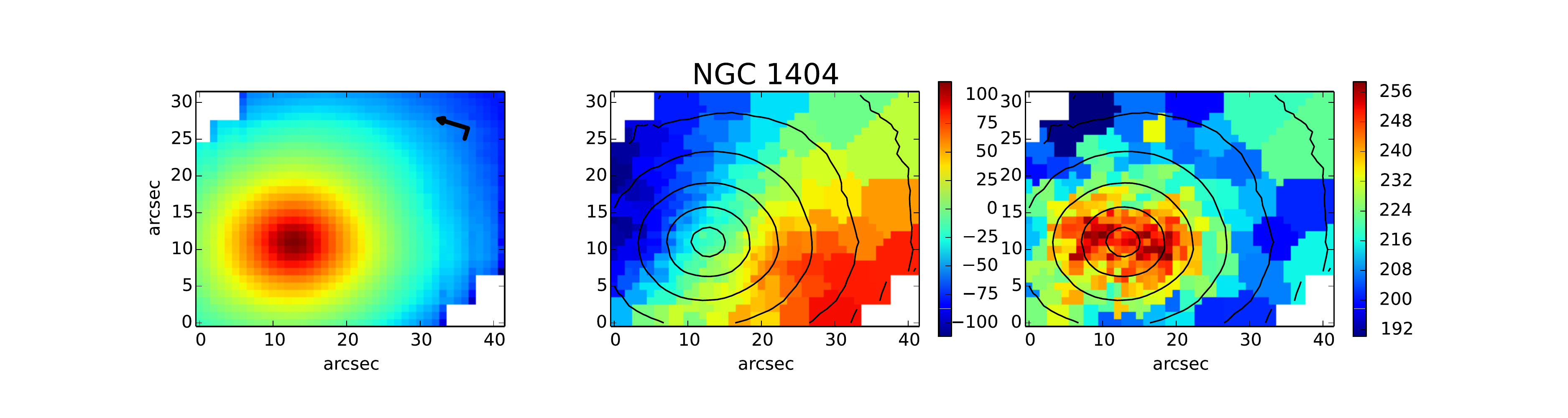}
\includegraphics[width=6in,clip,trim= 100 40 100 50]{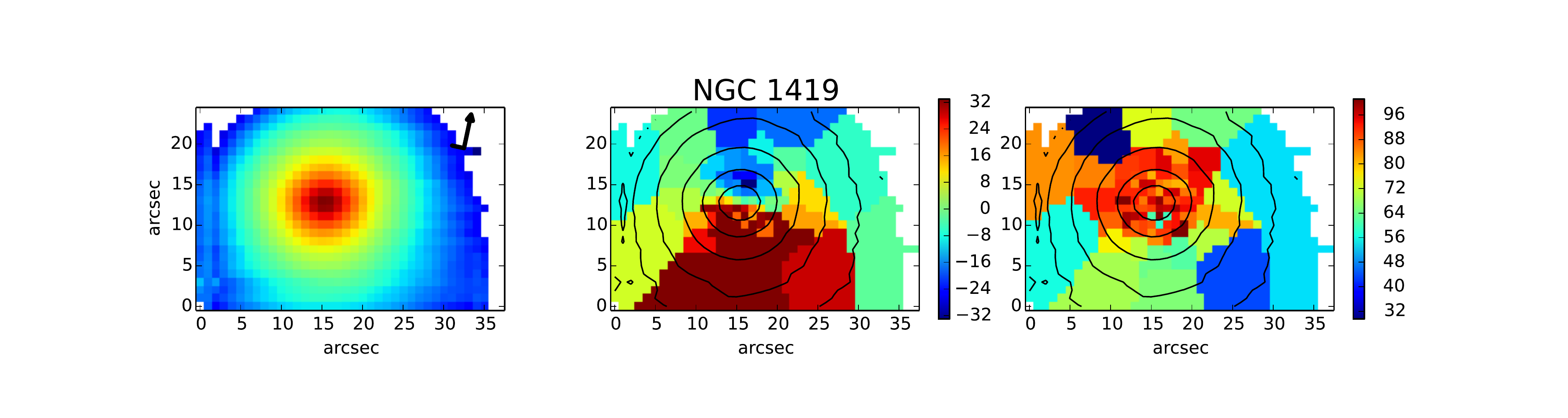}
\includegraphics[width=6in,clip,trim= 100 30 100 40]{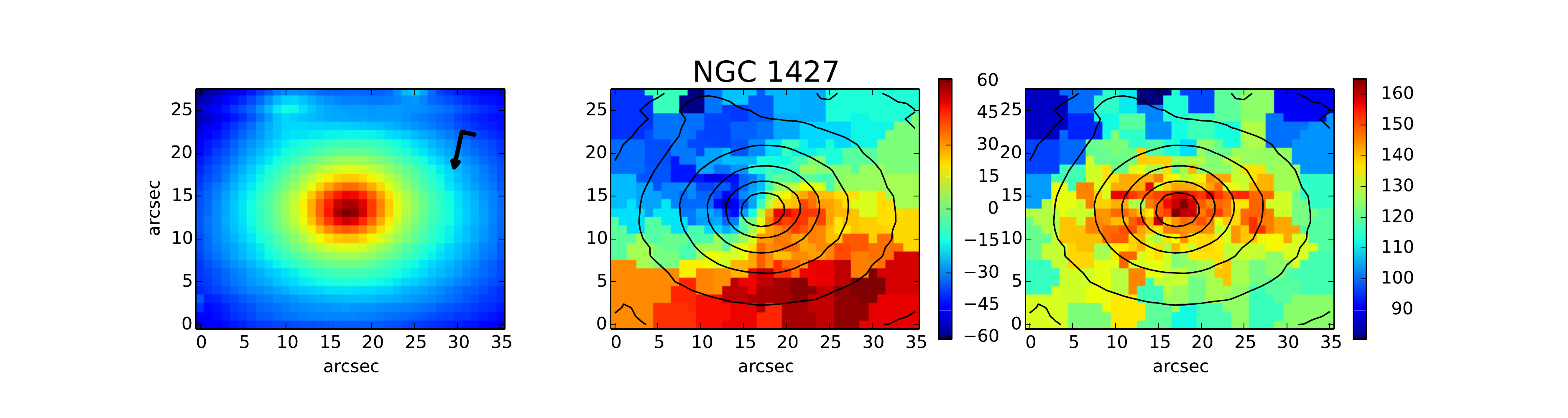}

Figure \ref{fig:figurea1} continued.
\end{figure*}

After determining the optimal stellar template from the high S/N central aperture spectrum, we then ran pPXF on the spatially binned spectra determined from the Voronoi binning technique. At this stage the stellar template is fixed to the optimal one, and only the LOSVD and additive polynomial are allowed to vary. This results in a measurement of the stellar velocity and stellar velocity dispersion for each independent spatial bin. Maps of the integrated flux (formed by collapsing each data cube in the spectral direction), stellar velocity and stellar velocity dispersion for the 10 objects in our sample for which we obtained WiFeS observations are presented in Figure \ref{fig:figurea1}.

\section{Supplementary data}
\label{sec:phot}

In addition to spatially resolved stellar kinematics several further pieces of information are required to construct the kinematic morphology--density relation for the Fornax Cluster. To classify galaxies into FRs and SRs both the ellipticity, $\epsilon$, and effective radius, R$_e$, of each galaxy must be determined. It is also necessary to characterise the local environmental density of each galaxy within the cluster. The measurement of these quantities, along with the data from which they were measured is described in the following subsections.

\subsection{Measurements of R$_e$ and $\epsilon_e$}
\label{sec:mge}

Part of the motivation for targeting the Fornax Cluster was the existence of high spatial resolution, high S/N, Hubble Space Telescope (HST) imaging for the majority of ETGs in the cluster. This imaging was obtained as part of the Advanced Camera for Surveys Fornax Cluster Survey \citep[ACSFCS,][]{Jordan:2007}. Of our 10 observed ETGs, 9 have ACS imaging in the F475W and F814W bands. For the remaining galaxy, NGC~1379, Wide Field Planetary Camera 2 (WFPC2) data is available in the F450W band.

For each galaxy we first constructed a Multi Gaussian Expansion (MGE) model \citep{Emsellem:1994} of its surface brightness profile from the HST F475W or F450W image using the method and software of \citet{Cappellari:2002} as detailed in \citet{Scott:2013}. The surface brightness profile is described as the sum of a set of two-dimensional Gaussians with varying normalisation, width and axial ratios. This formalism is extremely flexible and has been shown by \citet{Scott:2013} to provide a good description of the surface brightness profiles of a wide variety of galaxy morphologies, as well as giving accurate total magnitudes. From the MGE model we then determine the effective radius of the object following \citet[their equaltion 11]{Cappellari:2013a}. The MGE model was first circularised (setting the axial ratios of each Gaussian component to 1 while preserving the peak surface brightness of each component), then the radius which encloses exactly half the total luminosity of the model was determined by interpolating over a grid of radial values.

We determined ellipticity and position angle profiles directly from the HST images using the {\sc idl} routine {\it find\_galaxy.pro} following \citet{Krajnovic:2011}. The ellipticity and position angle at a given radius are determined from the second moments of the luminosity distribution of all connected pixels above a given flux level. The values of the ellipticity and the position angle determined at the isophote that encloses an area equal to $\pi$R$_e^2$ are distinguished with a subscript $e$: $\epsilon_e$ and PA$_e$. For NGC~1382, which has a prominent bar, we adopted a representative $\epsilon$ and PA measured at large radius, beyond the extent of the bar.  For each object, $\epsilon_\mathrm{e}$ is given in Table \ref{tab:table1}.

\subsection{Characterisation of environment}
We follow exactly the method of characterising the local projected environmental density presented in \citet{Cappellari:2011b}. We elect to use the $\Sigma_3$ density estimator, defined as the cylinder whose radius encloses the three nearest neighbouring galaxies and whose depth encloses all cluster members along the line of sight. \citet{Cappellari:2011b} showed this environmental measure to be the most sensitive to variations of kinematic morphology within the ATLAS$^\mathrm{3D}$ survey. Local projected environmental densities were calculated from an expanded parent sample, again drawn from the 2MASS Extended Source Catalogue, but including all galaxies within 10 degrees of the cluster centre to avoid edge effects. The same apparent magnitude limit, m$_{K_s} < 10.1$ was used to calculate $\Sigma_3$. Recession velocities for all objects were drawn from the 2MASS Redshift Survey \citep[2MRS,][]{Huchra:2012}. After applying the apparent magnitude and cluster-centric radius selection, we found that all galaxies had recession velocities consistent with the cluster recession velocity of $\sim 1400$ km\ s$^{-1}$, and therefore did not apply an additional velocity selection. The $\Sigma_3$ measurements for the early-type galaxies in our sample are given in Table \ref{tab:table1}. \citet{Muldrew:2012} show that nearest-neighbour projected environmental measures such as the one we use here are most suitable for probing the internal structure of a halo.

\section{Results}
\label{sec:results}

\begin{figure}
\includegraphics[width=3.25in]{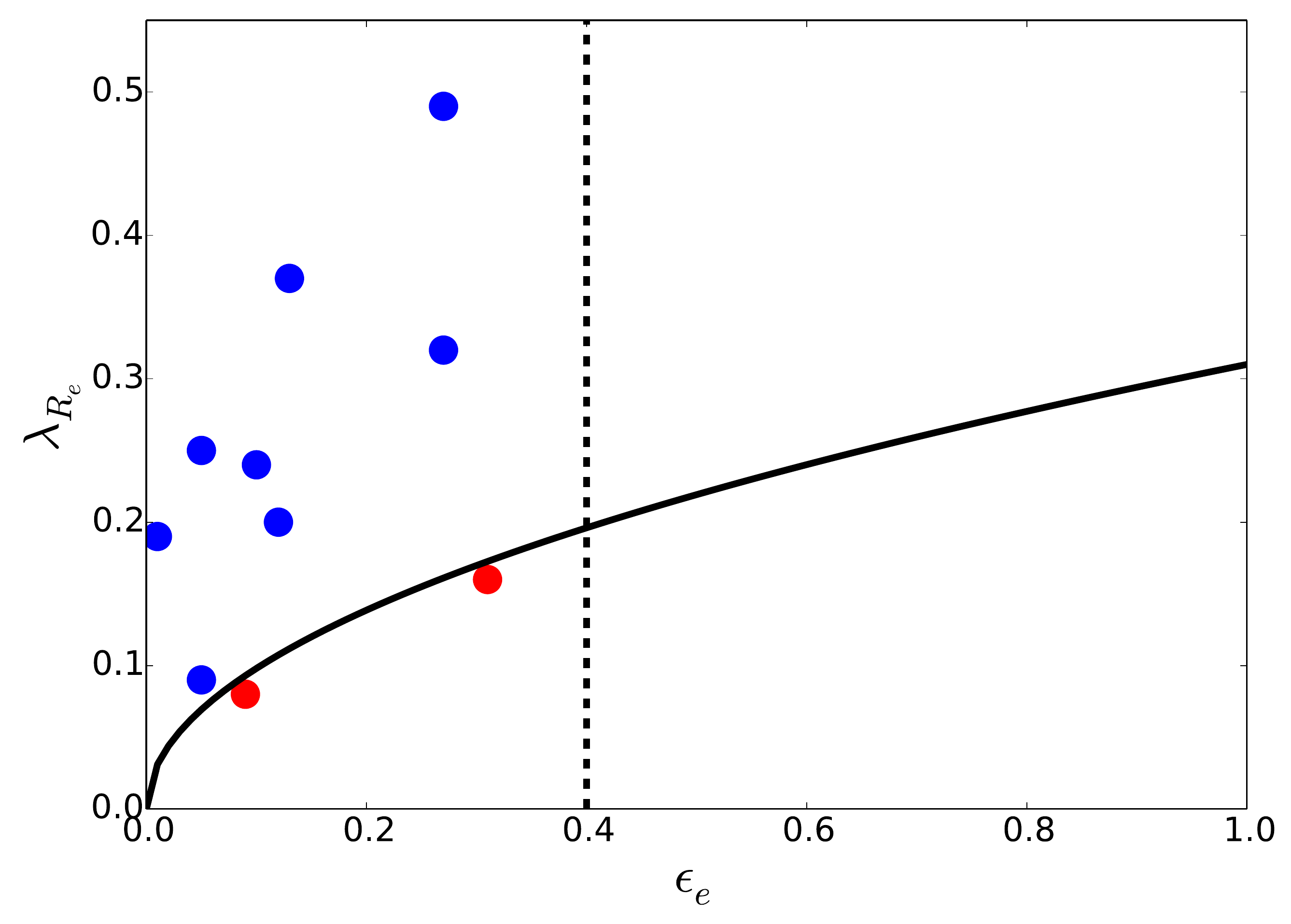}
\caption{$\lambda_{R_e}$ vs. $\epsilon_e$ for the 10 galaxies with WiFeS stellar kinematics in our sample. The solid black curve shows the division between slow- and fast-rotators from \citet{Emsellem:2011}. Galaxies above the line are FRs (blue points), galaxies below the line are SRs (red points). The dashed line indicates the ellipticity selection limit for our IFS targets. For $\epsilon _e< 0.4$ we find the same distribution of $\lambda_{R_e}$ values as was found by \citet{Emsellem:2011}.}
\label{fig:figure4}
\end{figure}

\subsection{Kinematic Classification}
Galaxies can be classified based on their kinematic morphology using two complementary approaches: by-eye classifications of the morphology of their velocity maps and a quantitative approach using the specific stellar angular momentum, $\lambda_R$ \citep{Emsellem:2007}. Where the data quality is sufficiently high (as in this work), the morphology of the velocity map is a more informative classification technique than that based on $\lambda_R$ alone.

On inspection of the velocity maps, the majority of galaxies show regular velocity fields with the rotation aligned along the photometric major axis and with significant rotation velocities. 7/10 galaxies have velocity fields whose morphology is consistent with being FRs. The exceptions to this are NGC~1399, NGC~1382 and NGC~1427. NGC~1399 shows no ordered rotation, consistent with its classification as a SR. NGC~1427 shows a kinematically decoupled component in the central 7". The central kinematically decoupled component is aligned with the photometric major axis, however outside this region the rotation is almost perpendicular to this axis. The kinematically decoupled component, along with the relatively low rotation velocities and high central dispersion confirm this galaxy as a SR. NGC~1382 shows a regular velocity field aligned with the photometric major axis of the galaxy, though the isophotes in the WiFeS field-of-view are dominated by a large bar component that is approximately perpendicular to the photometric position angle in the inner and outer parts of the galaxy. Despite being ordered, the magnitude of rotation in this object is extremely small, only $\pm 15 $km s$^{-1}$. Ordered rotation of such small amplitude is unusual in the ATLAS$^\mathrm{3D}$ sample. Further complicating the picture, the strong bar in the galaxy suggests that it contains a significant stellar disk, though this is not apparent from the kinematics. This suggests the galaxy is a disk-dominated system seen close to face on and, therefore, a FR. We also note here that, while it does not alter its kinematic classification, the velocity dispersion map of NGC~1404 shows evidence for two separated peaks, indicative of a small counter-rotating component, that is not apparent in the velocity map alone \citep[see][]{Franx:1989}. This gives 2/10 galaxies classified as SRs based on a visual inspection of their velocity maps. 

\begin{figure*}
\includegraphics[width=6.8in]{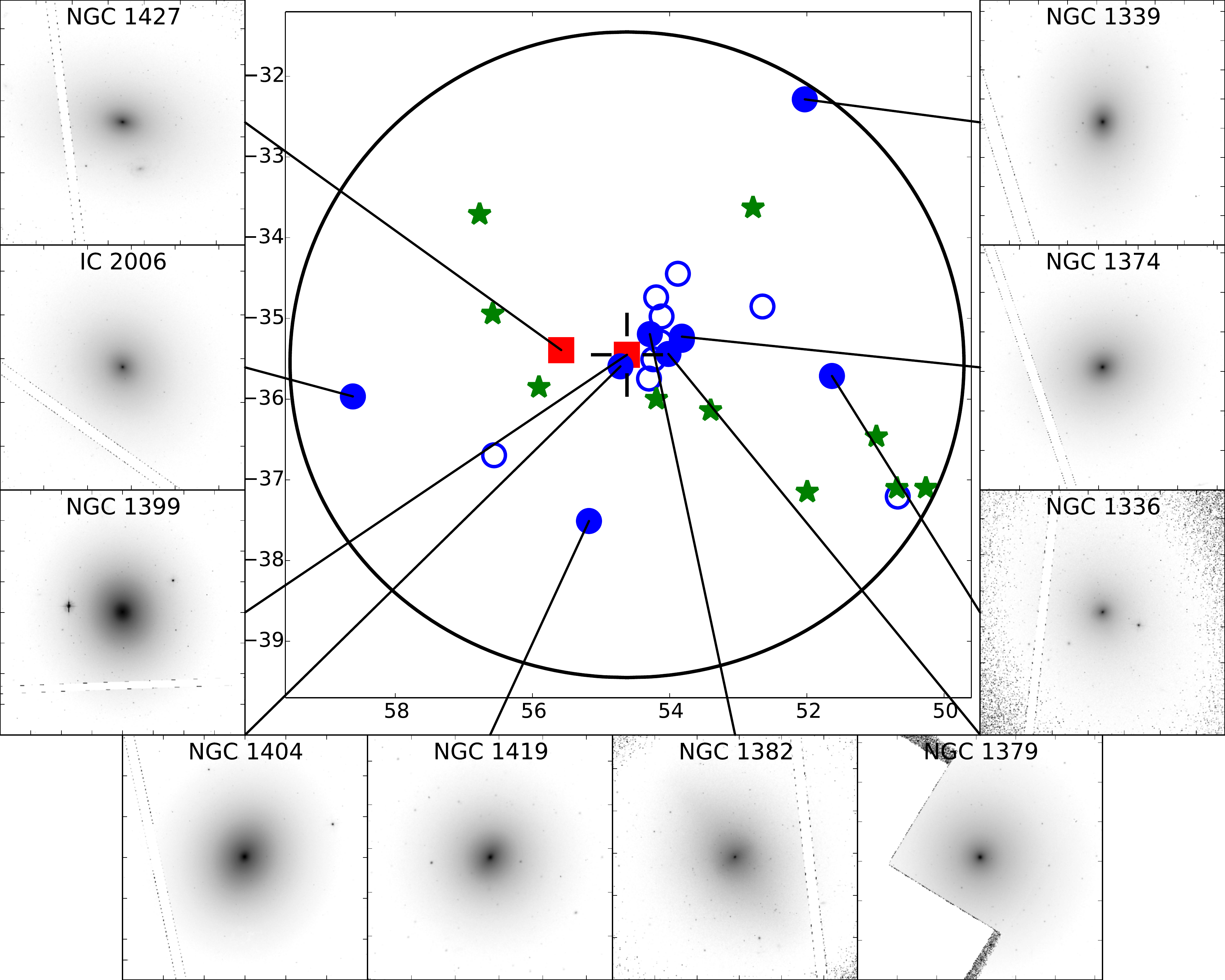}
\caption{Position of FRs (blue circles), SRs (red squares) and spirals (green stars) brighter than M$_K = -21.5$ mag within the Fornax cluster. The open blue circles show likely FRs based on their apparent flattening or literature kinematics, for which we did not obtain WiFeS stellar kinematics. The solid black line indicates a radius of 4 degrees, the Virial radius of the cluster. The two SRs are found towards the centre of the cluster, though not in the highest density regions, with spirals tending to be found on the outskirts of the cluster. There is a local over-density centred on NGC~1316 in the south west, which is thought to indicate an in-falling group. The outer panels show the HST images used in the MGE modelling described in Section \ref{sec:mge}. Each image is 2 R$_e$ across. The main cluster map and all sub-images are oriented north up, east left.}
\label{fig:figure5}
\end{figure*}

A consistent picture arises from a classification based on the specific stellar angular momentum within one R$_e$, $\lambda_{R_e}$. This is defined as:
\begin{equation}
\lambda_R(R_e) = \lambda_{R_e} = \frac{\Sigma_{i=1}^N F_i R_i V_i}{\Sigma_{i=1}^N F_i R_i \sqrt{V_i^2 + \sigma_i^2}},
\end{equation}
where $F_i$, $R_i$, $V_i$ and $\sigma_i$ are the flux, radius, velocity and velocity dispersion of the $i$th bin, and the sum is over all bins within an ellipse of area $\pi R_e^2$ and ellipticity $\epsilon_e$. \citet{Emsellem:2011} defined SRs as those galaxies for which $\lambda_{R_e}/\sqrt{\epsilon_e} < 0.31$ (the dependency on $\epsilon_e$ accounts for the influence of the inclination of the system). This value was chosen to reflect the by-eye classification of the kinematic maps carried out in that survey \citep{Krajnovic:2011}. While the $\lambda_R$-based approach is necessary for lower quality data, given the high S/N data presented here we make use of both approaches.

In Figure \ref{fig:figure4} we show the $\lambda_{R_e}$ vs. $\epsilon_e$ diagram for the 10 galaxies observed with WiFeS. The solid line indicates the division between FRs and SRs from \citet{Emsellem:2011}. $\lambda_{R_e}$ for these galaxies is given in Table \ref{tab:table1}. 2/10 galaxies fall below the SR line; these are NGC~1399 and NGC~1427. The remaining 8 galaxies all have values of $\lambda_{R_e}$ consistent with being FRs. This is consistent with the classification based on a visual inspection of the velocity maps.

\subsubsection{Classification of galaxies without WiFeS stellar kinematics}
\label{sec:without}

As noted previously, we were only able to obtain WiFeS observations for 10/20 ETGs in the Fornax Cluster. When scheduling our observations we prioritised those galaxies with ellipticities rounder than $\sim 0.4$. Of the 10 galaxies not observed with WiFeS, 7 have $\epsilon > 0.4$. In the ATLAS$^\mathrm{3D}$ sample of 260 ETGs, \citet{Emsellem:2011} found only one SR with $\epsilon > 0.4$, and this object, NGC~4550, is the prototypical example of a counter-rotating disc system and therefore not a true SR. Among the $\sim 120$ ATLAS$^\mathrm{3D}$ galaxies with $\epsilon > 0.4$ there are effectively no SRs, allowing us to confidently classify our 7 targets with $\epsilon > 0.4$ and no WiFeS stellar kinematics as FRs.

Of the remaining three galaxies without WiFeS observations, NGC~1351 and NGC~1316 have good-quality stellar kinematic measurements determined from long-slit spectroscopy \citep[from][respectively]{DOnofrio:1995,Bedregal:2006}. Both galaxies show rotation curves along their photometric major axes fully consistent with being FRs. While integral field spectroscopy is preferred to determine the morphology of the velocity map, long-slit observations that detect high rotation velocities along the major axis combined with low velocity dispersions can be used to make preliminary kinematic classifications. The remaining galaxy, NGC~1387 has a clear, strong bar, a feature that is only present in disc systems. Therefore, despite this galaxy being round and having no literature kinematic measurements, the presence of a bar is sufficient to classify this system as a likely FR. We indicate the early-type galaxies without WiFeS stellar kinematics (and therefore with less certain kinematic classifications) with open symbols in Figure \ref{fig:figure5}.

\subsubsection{Non-kinematic classification of early-type galaxies}

The classification of galaxies into SRs and FRs is closely related, but not identical to, the division of galaxies into those with flattened (or core-S\'ersic) and cuspy (or S\'ersic) central light profiles \citep[see][for an overview of the classification scheme]{Graham:2003a}. \citet{Krajnovic:2013} carried out a detailed comparison of the two classification schemes, finding that while in general SRs are typically core galaxies, there is a significant population of core FRs and a potentially unidentified population of SRs without cores. \citet{Glass:2011} examined the central surface brightness profiles of Fornax cluster galaxies, providing inner profile slopes which can be used to identify a core galaxy. Using their measurements of the $\Delta_3$ parameter which quantifies the inner light excess or deficit, we classify 2 galaxies, NGC~1399 and NGC~1316 as core-S\'ersic galaxies, with the remaining 18 early-type galaxies being S\'ersic galaxies \citep[though][report that the core classification for NGC~1316 remains uncertain]{Graham:2013}. Given the small sample size we cannot draw any significant conclusions on the connection between core type and kinematic type. We do note that NGC~1427 is an example of the core-less SR class that \citet{Krajnovic:2013} postulated.

\subsection{The Kinematic Morphology -- Density Relation within Fornax}
\label{sec:within_fornax}

\begin{figure}
\includegraphics[width=3.25in]{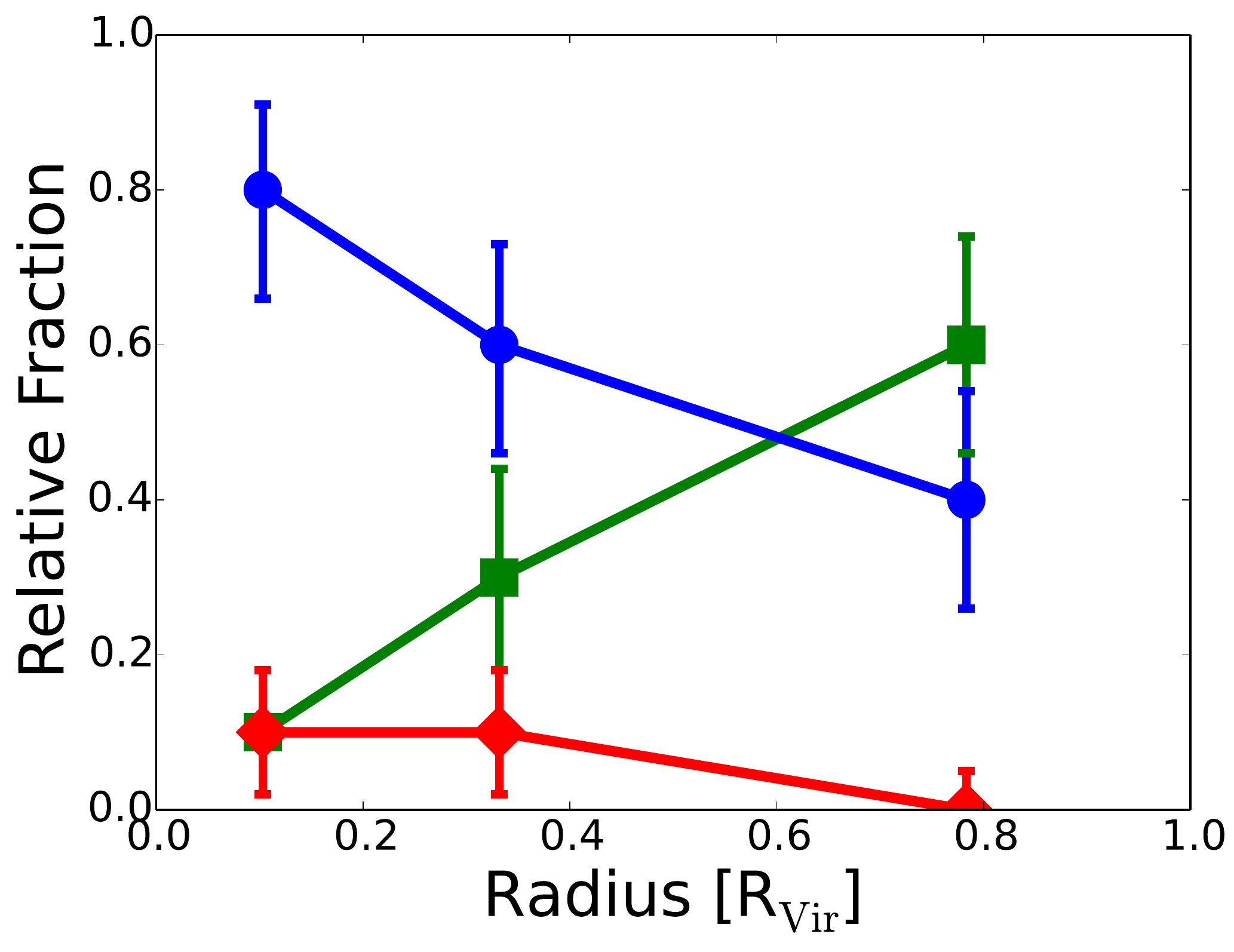}
\caption{Fraction of SRs (red), FRs (blue) and spirals (green) brighter than M$_K = -21.5$ mag as a function of projected radius from the centre of the Fornax cluster, normalised to the Virial radius of the cluster. There are 10 galaxies in each of the three bins. SRs are concentrated towards the centre of the cluster, with spirals found predominantly in the cluster outskirts.}
\label{fig:figure6}
\end{figure}

Of the 30 galaxies in the Fornax cluster with M$_K < -21.5$ mag, we find that 10 are spirals (based on the presence of spiral arms in imaging data), 2 are SRs (based on their kinematic morphology and $\lambda_{R_e}$) and the remaining 18 are FRs. The relative fraction of spirals, SRs and FRs are 33$^{+9}_{-8}$ per cent, 7$^{+4}_{-6}$ per cent and 60$^{+8}_{-9}$ per cent respectively. In Figure \ref{fig:figure5} we show the on-sky position of SRs (red squares), FRs (blue circles) and spirals (green stars) within the Fornax cluster, with the brightest cluster galaxy, NGC~1399 at the centre. The two SRs are found close to the centre of the cluster, with FRs distributed throughout the cluster and spirals generally located towards the outskirts of the cluster. This impression is quantified in Figure \ref{fig:figure6}, where we show the relative fraction of SRs (red), FRs (blue) and spirals (green) as a function of projected radius from the cluster centre. The fraction of SRs, relative to the total number of galaxies of all kinematic types in each radial bin, increases to 16$^{+11}_{-8}$ per cent at small values of the cluster-centric radius and is zero outside 0.2 Virial radii. In contrast there is a dearth of spirals in the inner radial bins, whereas $\sim 50$ per cent of galaxies in the outer part of the cluster are spirals. FRs are found at all cluster-centric radii, making up the dominant population within 0.4 Virial radii, and outside this radius being comparable in number to spiral galaxies.

\begin{figure}
\includegraphics[width=3.25in]{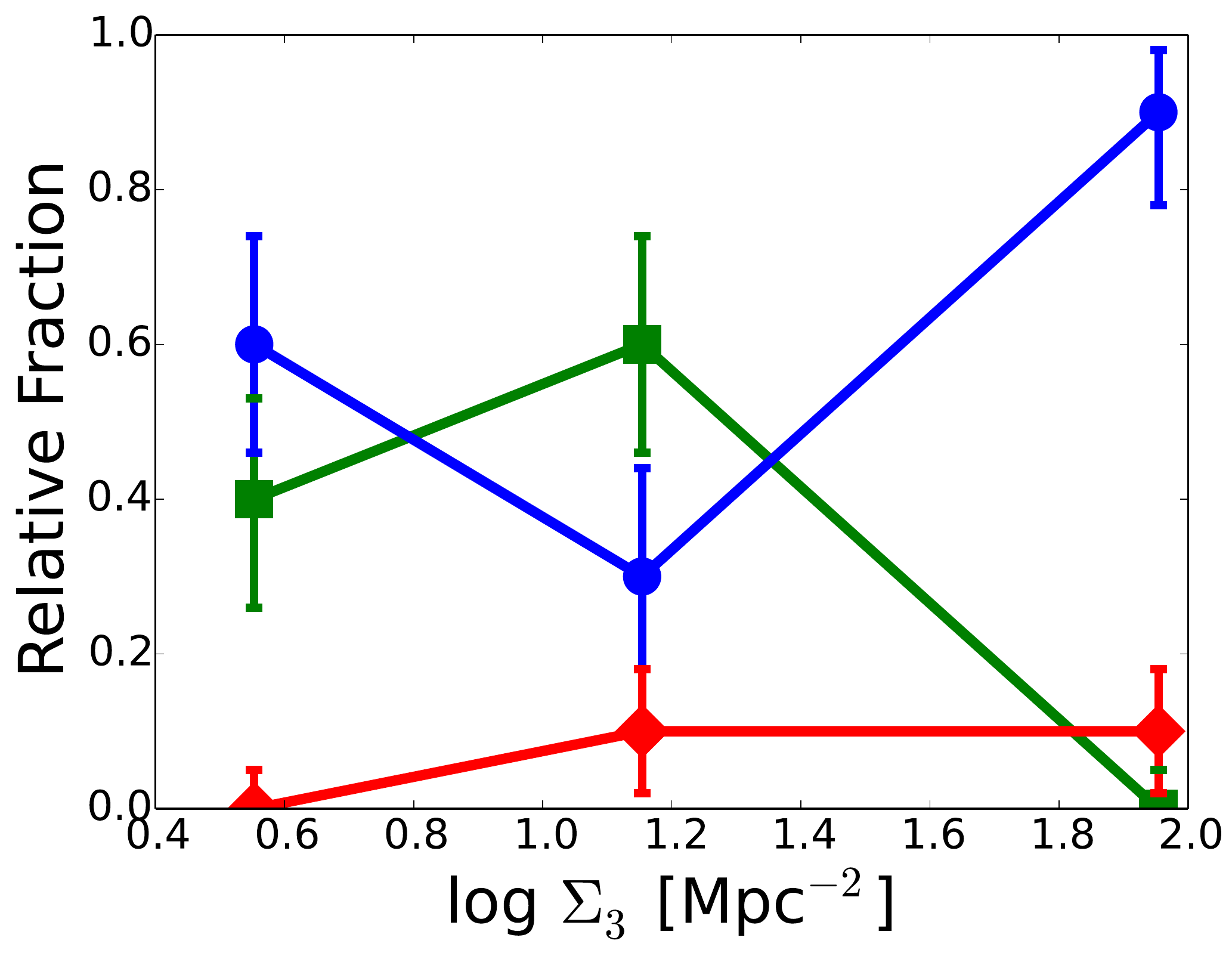}
\caption{Fraction of SRs (red), FRs (blue) and spirals (green) brighter than M$_K = -21.5$ mag as a function of local projected environmental density, $\Sigma_3$ for the Fornax cluster. SRs are more common at higher projected environmental densities and spirals are predominantly found at the lowest densities, while FRs are found throughout the cluster. There are 10 galaxies in each of the three bins of local projected environmental density.}
\label{fig:figure7}
\end{figure}

We also examine in Figure \ref{fig:figure7} how the relative fractions of each kinematic type vary with $\Sigma_3$, the local projected environmental surface density. Curiously, the peak in $\Sigma_3$ is offset from the brightest cluster galaxy (and commonly assumed cluster centre) NGC~1399, though when the three dimensional positions of cluster members are considered it is possible that NGC~1399 sits at the centre of the cluster potential. While SRs are not found at the lowest projected environmental densities and spirals avoid the highest densities, at intermediate densities all kinematic types are present. This may be due to the projected structure of the Fornax cluster, resulting in the peak in $\Sigma_3$ being offset from the cluster centre. Alternatively, it may indicate that kinematic type correlates more closely with cluster-centric radius than local projected environmental density.

\subsection{The Kinematic Morphology -- Density Relation across all environments}

\begin{figure}
\includegraphics[width=3.25in,clip,trim = 0 0 0 0]{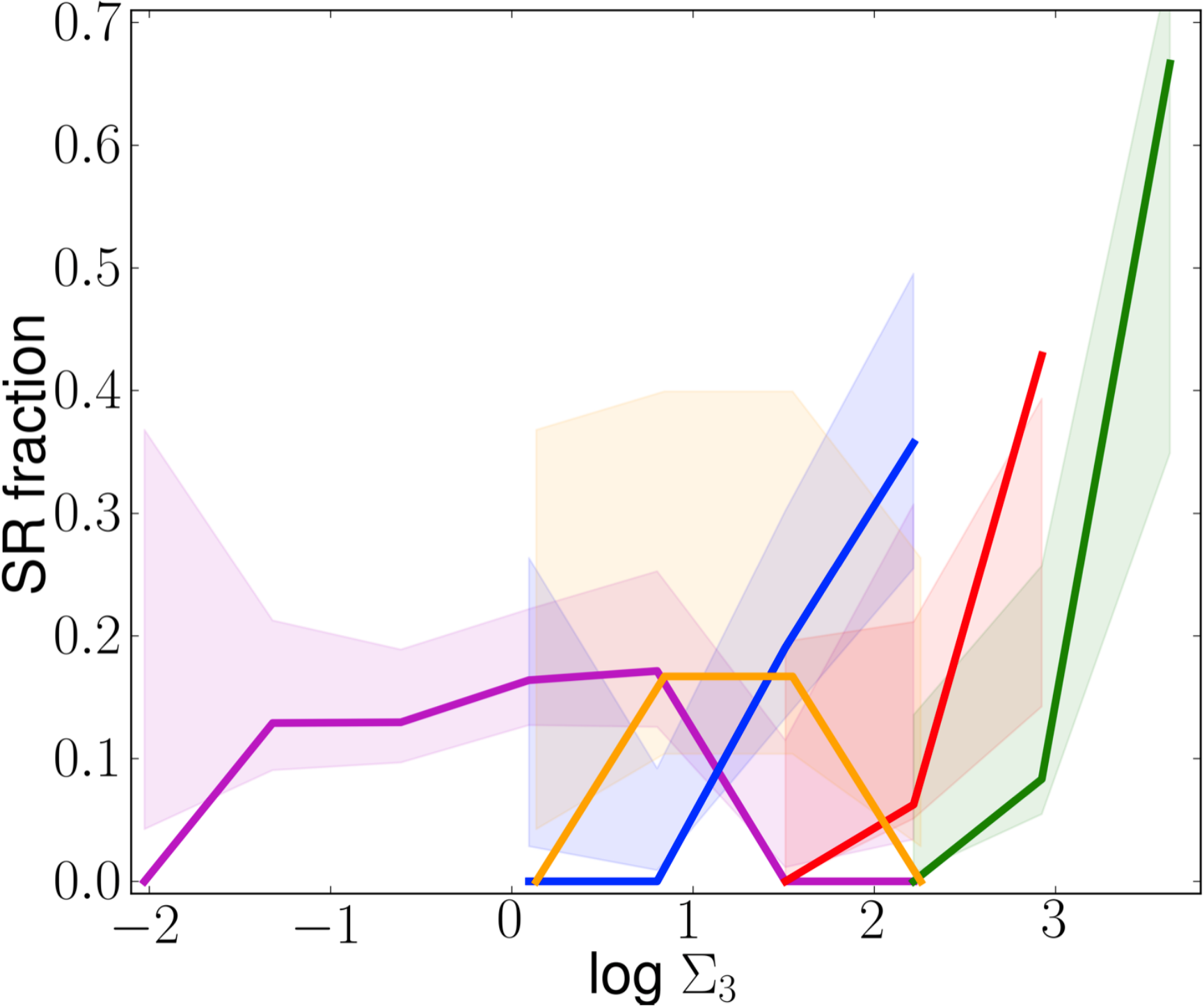}
\caption{Slow rotator fraction N$_\mathrm{SR}$/(N$_\mathrm{SR}$ + N$_\mathrm{FR}$) vs. local projected environmental density for the Fornax cluster (yellow line), the Virgo cluster \citep[blue line]{Cappellari:2011b}, the Coma cluster \citep[red line]{Houghton:2013} and the Abell~1689 cluster \citep[green line]{DEugenio:2013}. The ATLAS$^\mathrm{3D}$ field sample is shown in magenta. Most clusters show the same trend of increasing SR fraction from the low-density outskirts to the high-density centre. The projected environmental density at which this increase occurs varies from cluster to cluster. This behaviour is weaker in the Fornax cluster than the more massive clusters, though this may be due to the small sample size available in any single cluster with the richness of Fornax.}
\label{fig:figure8}
\end{figure}

As described in Section \ref{sec:intro}, similar integral-field studies of early-type galaxies have now been undertaken spanning a broad range of environments. By combining the data from these surveys we can examine how the SR fraction varies as a function of environment from the field to the centre of massive clusters like Coma and Abell~1689. We now consider the fraction of SRs with respect to the total number of early-type galaxies only: N$_\mathrm{SR}$/(N$_\mathrm{SR}$ + N$_\mathrm{FR}$). The most comprehensive sample examined so far was that of \citet{Houghton:2013}, who combined datasets covering the local field, the Virgo cluster \citep[both taken from][]{Cappellari:2011b}, the Coma cluster \citep{Scott:2012,Houghton:2013} and Abell~1689 \citep{DEugenio:2013}. All three studies used the same absolute magnitude limit of M$_K < -21.5$ mag, as chosen for this work. Here we simply reproduce in Figure \ref{fig:figure8} the analysis of \citet{Houghton:2013}, with the addition of the data for the Fornax cluster from this work. 

The observed SR fraction for the bright early-type galaxies in the Fornax cluster is 2/20, however, as in \citet{Houghton:2013}, we would like to use our observations to infer the probability distribution of SRs. As discussed extensively in \citet[their section 3.3.2]{Houghton:2013} the appropriate statistical analysis to derive the probability distribution of the SR fraction in the Fornax cluster from our observations is the binomial distribution. The binomial distribution determines the probability of $k$ successes, given $n$ trials and a probability $p$ of success, P($k|n,p$). For our analysis, $k$ is the observed number of SRs, $n$ the total number of ETGs, and $p$ the SR fraction of Fornax which we wish to determine. Applying this analysis to our results, we find a SR fraction of $13^{+8}_{-6}$ per cent in the Fornax cluster.

\section{Discussion}
\label{sec:disc}

By combining the results presented in this study with the kinematic morphology -- density relations presented in \citet{Houghton:2013} and the conclusions of \citet{Cappellari:2011b} we can construct a picture for the formation of early-type galaxies over a broad range of environments. Three main points arise from the results of these previous works and the new combined analysis presented here:
\begin{enumerate}
\item The late-type galaxy fraction decreases smoothly from low to high local environmental densities, with a corresponding increase in the early-type galaxy fraction. 
\item Approximately 15 per cent of early-type galaxies brighter than M$_K = -21.5$ mag in all studies to date are found to be slow rotators. This was first identified by \citet{DEugenio:2013} and \citet{Houghton:2013} and is confirmed in this work.
\item Within a halo, slow rotators are strongly concentrated towards small radii and high local projected environmental densities, with a corresponding deficit of slow rotators in the outskirts.
\end{enumerate}

\subsection{The Fornax Cluster versus more massive clusters}

Figure \ref{fig:figure8} summarises the variation of the SR fraction for a range of cluster environments. The Fornax cluster is the lowest mass cluster studied to date, but it is more compact than the more massive clusters and therefore covers a similar range of local projected environmental densities. The fraction of galaxies within the Fornax cluster of each kinematic type (spiral, FR, SR) is approximately consistent with the Coma cluster, but with a lower early-type fraction than the Virgo cluster. In all the environments explored to date, $\sim 15$ per cent of the early-type galaxy population brighter than M$_K = -21.5$ mag are SRs. However, as noted in Sec. \ref{sec:within_fornax}, the SR fraction varies strongly as a function of cluster-centric radius and local projected environmental density. Because the current literature samples probe quite different ranges in cluster-centric radius this constancy may be coincidental, though further observations covering a more homogeneous environmental range are required to assess this.

The outskirts of the Fornax cluster exhibit the same deficiency of SRs as the more massive Virgo, Coma and Abell~1689 clusters, with this region being dominated by early-type FR galaxies. The SR fraction in Fornax does increase towards the centre of the cluster, though the degree of segregation is less pronounced than in the more massive clusters, despite the similar local projected environmental densities spanned. This may be driven by projection effects, which cause the peak in local (projected) environmental density to be offset from the centre of the Fornax cluster. It is also important to note that similar projected densities do not necessarily represent similar physical densities, due to different distributions along the line-of-sight. As the Virgo and Fornax clusters have similar physical depths along the line-of-sight \citep{Blakeslee:2009} the projected and physical environmental densities are likely comparable. For the richer environments, which are likely more extended along the line-of-sight, the projected density is likely over-estimated with respect to Fornax and Virgo, however it is challenging to quantify this effect with existing data.

An alternative explanation presents itself when comparing the SR fraction vs. $\Sigma_3$ profile for the Fornax cluster (yellow line) to that for the ATLAS$^\mathrm{3D}$ field sample (magenta line) in Figure \ref{fig:figure8}. This may suggest that whatever process segregates SRs in higher density environments may not be effective in lower mass halos like the Fornax cluster. Due to the limited number of massive galaxies found in a given cluster of the same mass as Fornax, the variation of the SR fraction with $\Sigma_3$ within the Fornax cluster remains highly uncertain. A larger sample of galaxies in clusters of a similar mass to Fornax is required to reduce the statistical uncertainties. 

\subsection{Formation and segregation of slow rotators}

A possible explanation for the observed segregation of SRs in cluster environments is dynamical friction \citep[as previously proposed by][]{Cappellari:2011b,DEugenio:2013,Houghton:2013}. Through interaction with the cluster potential, more massive galaxies lose kinetic energy and are driven to the cluster centre. As SRs are, on average, more massive than FRs \citep[for high- and intermediate-mass galaxies with M$_* \sim 6 \times 10^9$ M$_\odot$:][]{Emsellem:2011} this mechanism would lead to a higher central concentration of SRs than FRs, as observed.

The efficiency of dynamical friction depends on the mass of a given galaxy, and is unaffected by kinematic type. Therefore at a fixed mass, SRs and FRs should show the same degree of segregation, with the increased segregation of SRs being a product of their differing mass distribution. We can test this scenario by constructing mass-matched samples of SRs and FRs and comparing their mean local environmental densities. We use the ATLAS$^\mathrm{3D}$ Virgo cluster sample for this purpose as: i) of the clusters with detailed kinematic type classifications the Virgo cluster has the largest observed sample, ii) \citet{Cappellari:2013a} provide dynamical masses for all early-type galaxies in that sample and iii) limiting the sample to a single cluster eliminates the confusion of different regions having the same local environmental density but being in different environments (e.g. outskirts, centre) with respect to their host halo. This gives a sample of 58 galaxies, of which 9 are SRs and the remaining 49 are FRs.

As an initial simple test, we derived the mean $\Sigma_3$ for all ETGs more massive than $1 \times 10^{11} \mathrm{M}_\odot$, which gives a sample of 5 SRs and 5 FRs. The mean mass of the SRs, $4.1 \times 10^{11} \mathrm{M}_\odot$ is higher than that for the FRs, $2.5 \times 10^{11} \mathrm{M}_\odot$, though only by 0.2 dex. However, the mean local environmental density of the SRs, $\log \Sigma_3 = 1.84 \pm 0.40$, is $\sim 1 \sigma$ higher than that of the FRs, $\log \Sigma_3 = 1.39 \pm 0.45$.

\begin{figure}
\includegraphics[width=3.25in]{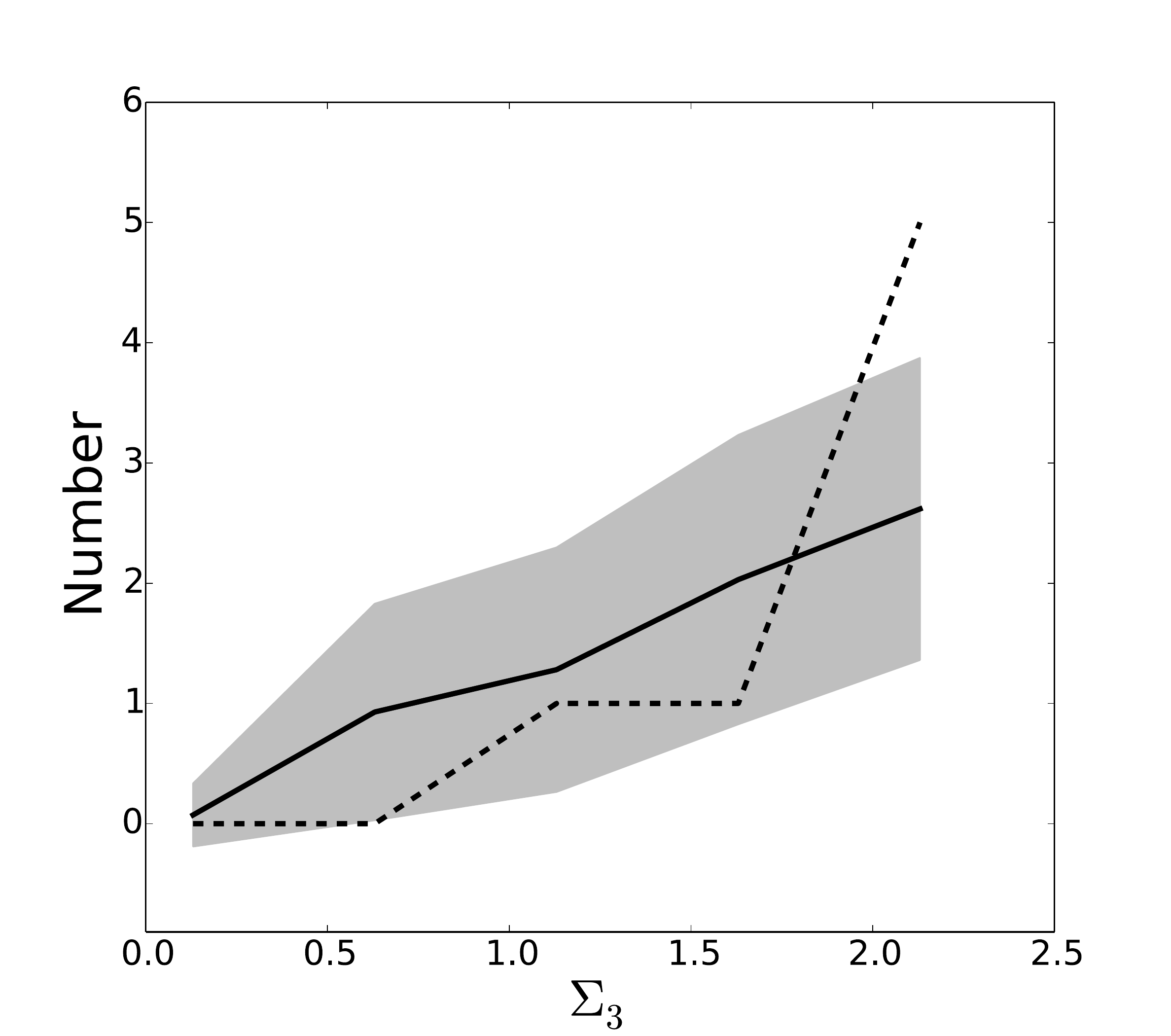}
\caption{Number of SRs (dashed line) and FRs (solid line) as a function of $\Sigma_3$ in the Virgo cluster. The shaded region indicates the $1-\sigma$ uncertainty on the number of FRs. The FR sample is matched in stellar mass to the SR sample as described in the text. Even at fixed mass, the SRs are found at much higher local densities than the FRs. This argues against dynamical friction leading to the observed segregation of SRs, assuming FRs and SRs have similar halo masses at a given stellar mass. Counter-rotating discs are excluded from the SR sample.}
\label{fig:figure9}
\end{figure}

We conducted an improved version of this test by constructing a random sample of FRs that is matched in its mass distribution to that of the SRs. By repeatedly selecting random samples that are mass-matched to the SR mass distribution we can improve our determination of the mean and standard deviation of the FR sample. While the mass-matching is not perfect due to the finite sample size, the mean mass of the FR samples selected in this way is only 0.1 dex smaller than that of the SR sample. Using this technique for the full sample of 7 Virgo SRs (excluding two counter-rotating discs which masquerade as SRs) we find: $\Sigma_3 = 1.91 \pm 0.36$, and for the FRs: $\Sigma_3 = 1.29 \pm 0.18$. This result is illustrated in alternative fashion in Figure \ref{fig:figure9}, where we show the number of SRs and the corresponding number of FRs in a mass-matched sample in bins of $\Sigma_3$. The dashed line of the SRs shows a significantly stronger trend to higher values of $\Sigma_3$ than the solid line of the FRs. At fixed galaxy mass, SRs occupy regions of significantly higher local environmental density than FRs. This shows that, assuming mass follows light, the SRs in Virgo cannot have formed together with the other galaxies in the cluster and reached the centre of the cluster by dynamical friction alone.

\citet{Cappellari:2013b} propose a more realistic scenario where SRs preferentially form in group environments as central galaxies, before joining the cluster environment.  They tend to remain centrals of their groups during the hierarchy of galaxy evolution, until they end up in the centre of massive clusters. This picture accounts for the hierarchical build-up of galaxy clusters predicted by galaxy formation models \citep[e.g.][]{DeLucia:2012}. In this scenario the SRs would typically be associated with a more massive halo than FRs of the same galaxy mass, increasing the efficiency of dynamical friction in driving these galaxies to the centre of their host clusters. While the idea that SRs form preferentially as central galaxies has not yet been rigorously observationally tested, the identification of a SR associated with a substructure within the Virgo cluster \citep[see][]{Cappellari:2011b} supports this arguement. SRs being preferentially central galaxies (and therefore being associated with a larger halo) has also been predicted by the semi-analytic simulation of \citet{Khochfar:2011}. Confirmation of this connection will require IFS surveys of statistically significant numbers of SRs in low-density environments. 

\section{Conclusions}

In this work we have presented integral-field data on a sample of 10 early-type galaxies brighter than M$_K = -21.5$ mag in the Fornax cluster. We derived spatially-resolved maps of the velocity and velocity dispersion for each galaxy, classifying them as either fast or slow rotators based on i) the morphology of their velocity maps and ii) their specific stellar angular momentum, $\lambda_R$. The remaining 10 early-type galaxies in the Fornax cluster that satisfy our selection criteria, but for which we were unable to obtain integral-field data, are highly likely to be FRs based on their apparent flattenings and previous long-slit spectroscopy.

From our kinematic classification we determine that 2/30 (7$^{+}_{-}$ per cent) galaxies in the Fornax cluster are slow rotators, 18/30 (60$^{+}_{-}$ per cent) are fast rotators and the remaining 10/30 (33$^{+}_{-}$ per cent) are spirals. Using the binomial distribution, we infer the fraction of early-type galaxies in the Fornax cluster that are slow rotators is $13^{+8}_{-6}$ per cent, consistent with the findings of \citet{Emsellem:2011} and \citet{Houghton:2013}. This fraction, observed for early-type galaxies brighter than M$_K = -21.5$, appears constant across all environments explored to date, though may only be a coincidence given the variation in cluster-centric radius probed by the different samples currently available.

As with previous studies of more massive clusters, we find that the slow rotators are strongly concentrated towards the cluster centre. Unlike in the Virgo, Coma and Abell~1689 clusters, the Fornax slow rotators are not strongly concentrated in the highest local-density region of the cluster --- although one of the two SRs is near the peak density, the other is offset. This difference is not very significant given the small number of galaxies in the cluster. Future studies involving statistically significant numbers of galaxies in low-mass clusters and groups will be required to identify whether the trends in slow rotator fraction described above are present in lower mass halos.  Such studies will be able to constrain the relative importance of different environmental transformation mechanisms that produce slow rotators as a function of a galaxy's local and global environment.

\section{Acknowledgements}

This research was conducted by the Australian Research Council Centre of Excellence for All-sky Astrophysics (CAASTRO), through project number CE110001020. NS and AWG acknowledge support of Australian Research Council funding through grants DP110103509 and FT110100263. RLD is grateful for support from the Australian Astronomical Observatory Distinguished Visitors programme, CAASTRO, and the University of Sydney during a sabbatical visit. RCWH was supported by the Science and Technology Facilities Council (STFC grant number ST/H002456/1). MC acknowledges support from a Royal Society University Research Fellowship. KAP thanks Christ Church, Oxford, for their hospitality whilst some of this work was being undertaken and is grateful to Oxford University for support during a sabbatical visit.

This publication makes use of data products from the Two Micron All Sky Survey, which is a joint project of the University of Massachusetts and the Infrared Processing and Analysis Center/California Institute of Technology, funded by the National Aeronautics and Space Administration and the National Science Foundation.

Based on observations made with the NASA/ESA Hubble Space Telescope, and obtained from the Hubble Legacy Archive, which is a collaboration between the Space Telescope Science Institute (STScI/NASA), the Space Telescope European Coordinating Facility (ST-ECF/ESA) and the Canadian Astronomy Data Centre (CADC/NRC/CSA).

This research has made use of the NASA/IPAC Extragalactic Database (NED) which is operated by the Jet Propulsion Laboratory, California Institute of Technology, under contract with the National Aeronautics and Space Administration. 

We acknowledge usage of the HyperLeda database (http://leda.univ-lyon1.fr).
\bibliographystyle{mn2e}
\bibliography{wifes_fornax}

\appendix
\section{Comparison of WiFeS kinematics to past studies}

\subsection{Comparison of WiFeS and ATLAS$^\mathrm{3D}$ kinematics for NGC~7710}

\begin{figure*}
\includegraphics[width=6.5in,clip,trim=0 0 35 20]{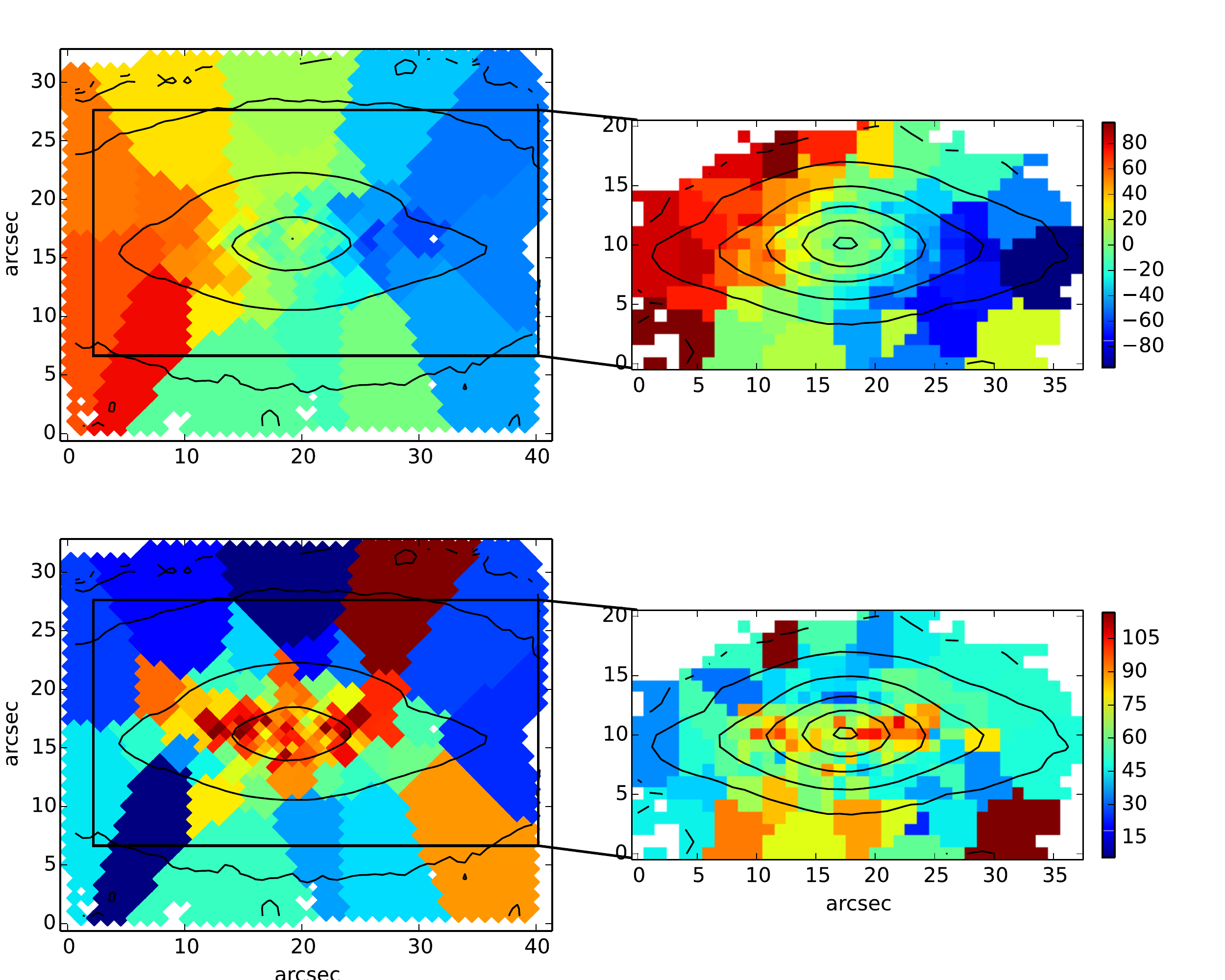}
\caption{Left panels: Stellar velocity and velocity dispersion maps for NGC~7710 from SAURON observations presented in \citet{Krajnovic:2011}. Right panels: Stellar velocity and velocity dispersion maps for NGC~7710 from the WiFeS observations presented in this work. NGC~7710 contains two counter-rotating disks of stars, clearly indicated by i) the double-peaked dispersion map that is symmetric about the galaxy centre and ii) the multiple changes in the sense of rotation along the major axis. The overlap between the WiFeS and SAURON field-of-view is indicated with the black rectangles in the left hand column.}
\label{fig:figure3}
\end{figure*}

In addition to 10 Fornax cluster galaxies, we observed one further galaxy, NGC ~7710, which had previously been observed by the ATLAS$^\mathrm{3D}$ survey. We extracted spatially resolved stellar kinematics for this object as described above, except that we adopted a lower target S/N of 10 for the Voronoi binning. This was done as NGC~7710 is at approximately double the redshift of our main sample, and therefore has considerably lower surface brightness in the outer parts of the WiFeS field-of-view compared to the majority of our targets. The lower target S/N was required to ensure reasonable spatial coverage of NGC~7710, while still maintaining high enough S/N to determine stellar velocities and velocity dispersions. Maps of the stellar velocity and stellar velocity dispersion for NGC~7710 are shown in the right hand column of Figure \ref{fig:figure2}, with the corresponding maps from the ATLAS$^\mathrm{3D}$ survey shown in the left column, taken from \citet{Cappellari:2011a}\footnote{The data are available at http://purl.org/atlas3d}. The ATLAS$^\mathrm{3D}$ maps were previously shown in \citet[Fig. C5 and C6]{Krajnovic:2011}.

NGC~7710 is classified by \citet{Krajnovic:2011} as a ``2--$\sigma$'' galaxy, an object showing two peaks in the velocity dispersion map, well separated and lying along the galaxy's photometric major axis. This object also shows multiple changes of direction in the sense of rotation along the major axis. These features in the velocity and velocity dispersion maps are classic signatures of a galaxy composed of two counter-rotating stellar disks. We chose NGC~7710 as our comparison object in order to determine how well we are able to detect the complex kinematic structures frequently found in early-type galaxies by the ATLAS$^\mathrm{3D}$ survey. 

The double-peaked structure along the major axis in the velocity dispersion map is clearly seen in both the WiFeS and ATLAS$^\mathrm{3D}$ maps. The magnitude of the dispersion in the peaks, $\sim 100 \mathrm{km s}^{-1}$ is the same in the WiFeS and ATLAS$^\mathrm{3D}$ data. The magnitude and sense of rotation in the outer parts of the WiFeS velocity map, $\sim \pm 80 \mathrm{km\ s}^{-1}$, is consistent with that observed at the same radius in the ATLAS$^\mathrm{3D}$ map. The WiFeS velocity map does not show the same reversal in the sense of rotation along the major axis as seen in the ATLAS$^\mathrm{3D}$ data, however the WiFeS velocity map does show an extended area of low velocity in the corresponding region. This suggests that the spatial resolution of the WiFeS map is not sufficient to resolve the fine structure of the velocity field in this case, though it is more than sufficient to resolve the global structure and slightly larger-scale details such as the separation of the two dispersion peaks. This is expected, given the larger spatial pixels of WiFeS (1'' compared to 0.8'' for SAURON) and the typically poorer seeing at the respective sites. Overall, we find excellent agreement between our own determination of the stellar kinematics of NGC~7710 from WiFeS observations and that presented by \citet{Krajnovic:2011} as part of the ATLAS$^\mathrm{3D}$ survey.

\subsection{Comparison of long-slit rotation curves extracted from WiFeS data to literature kinematics}

\begin{figure*}
\includegraphics[width=2.3in,clip,trim = 0 20 50 50]{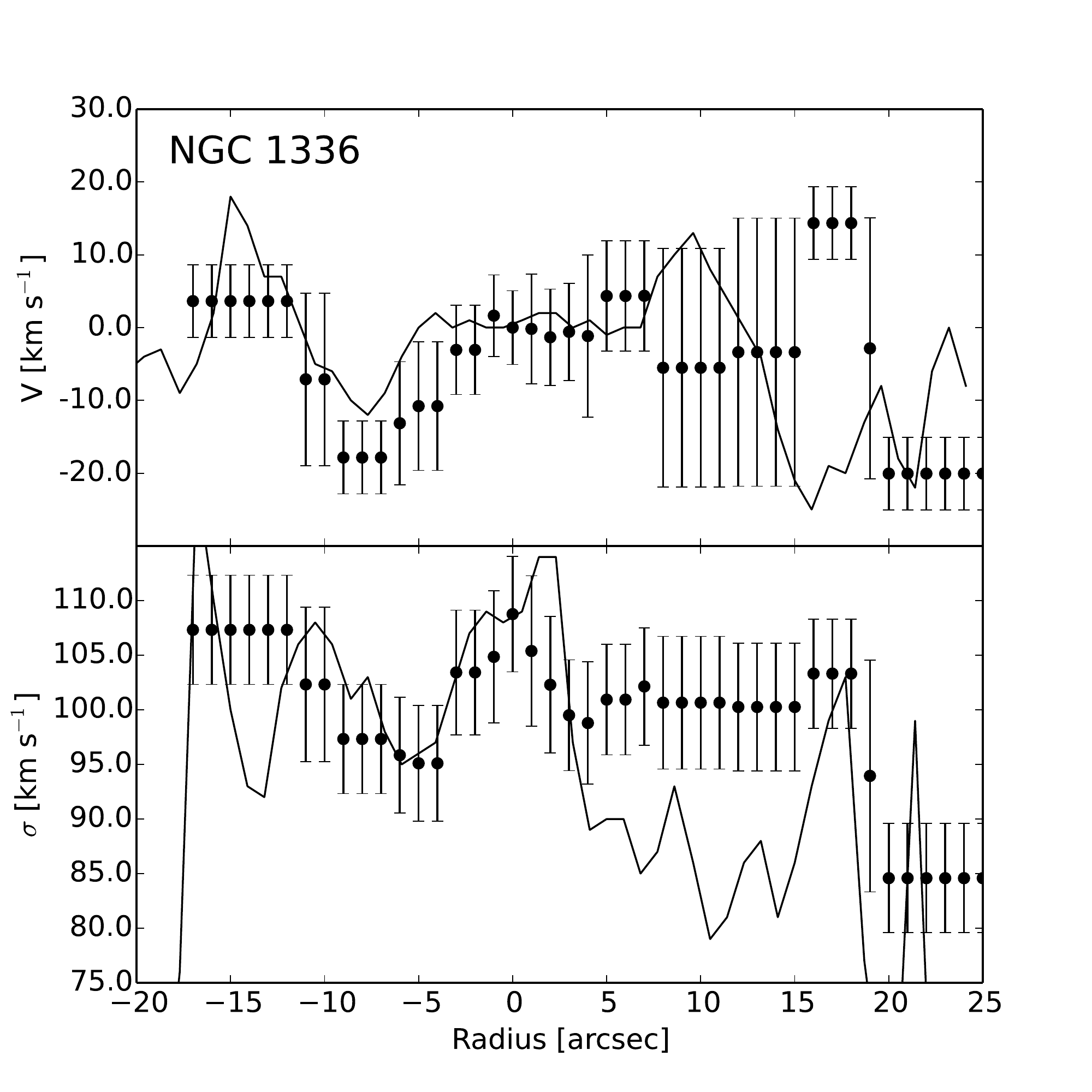}
\includegraphics[width=2.3in,clip,trim = 0 20 50 50]{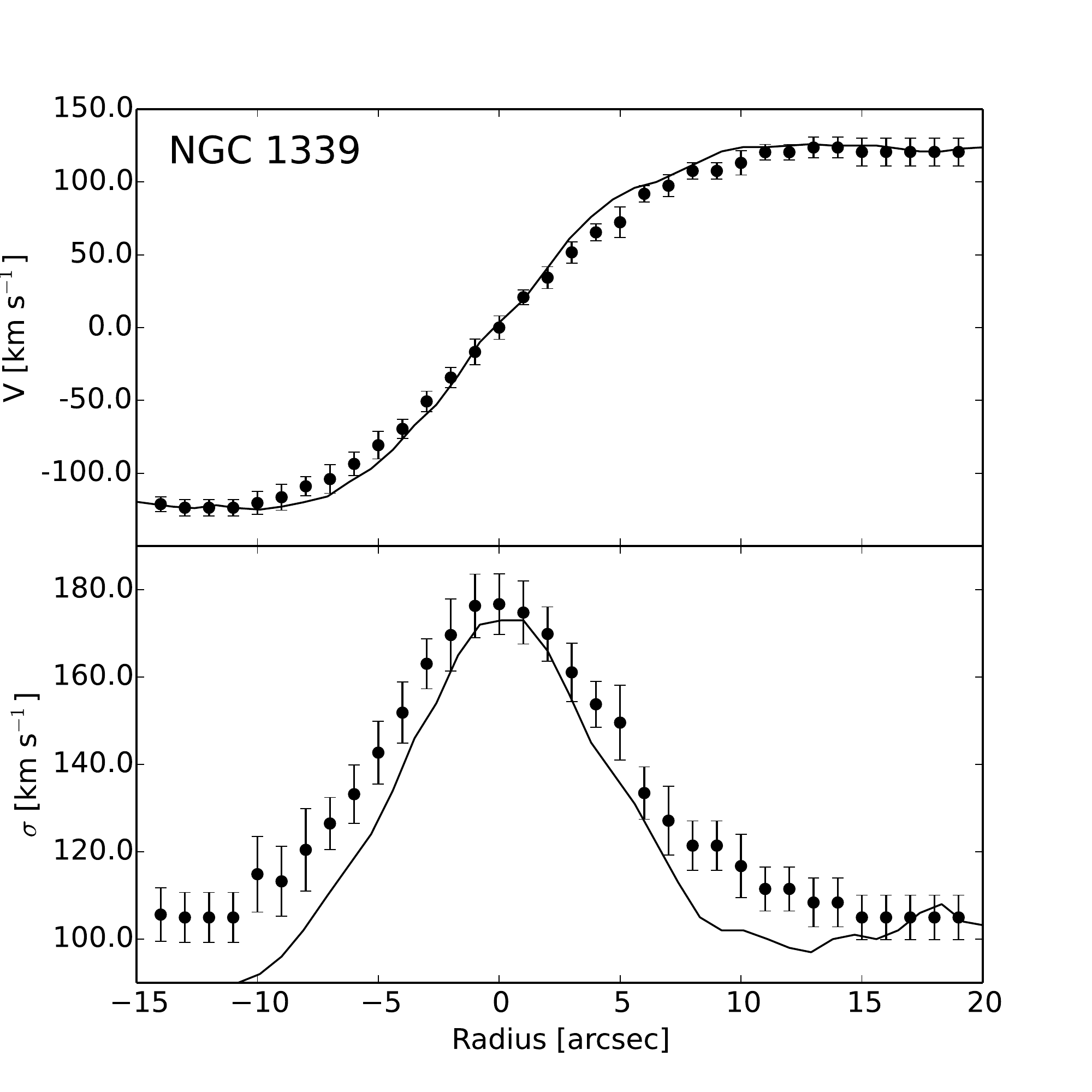}
\includegraphics[width=2.3in,clip,trim = 0 20 50 50]{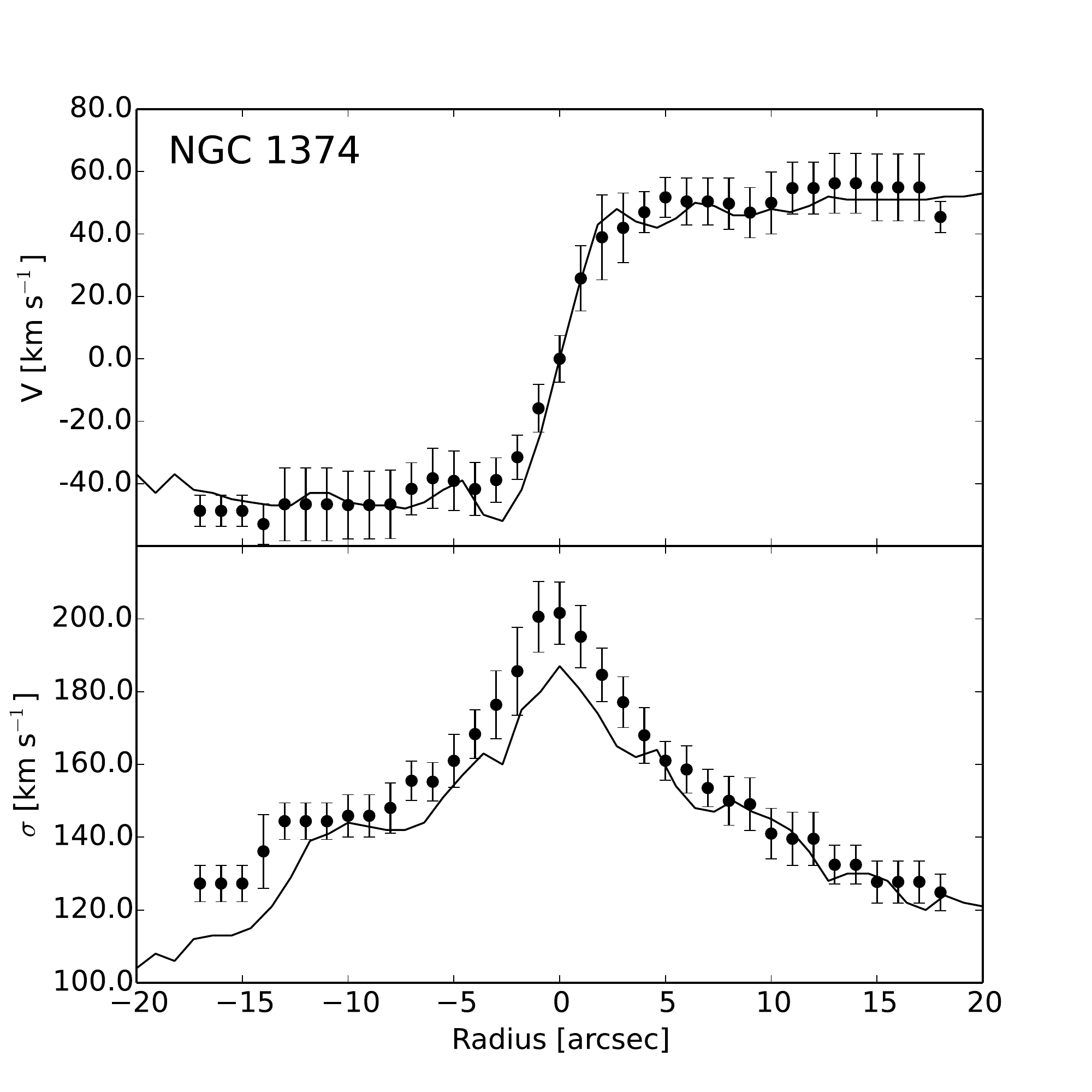}
\includegraphics[width=2.3in,clip,trim = 0 20 50 50]{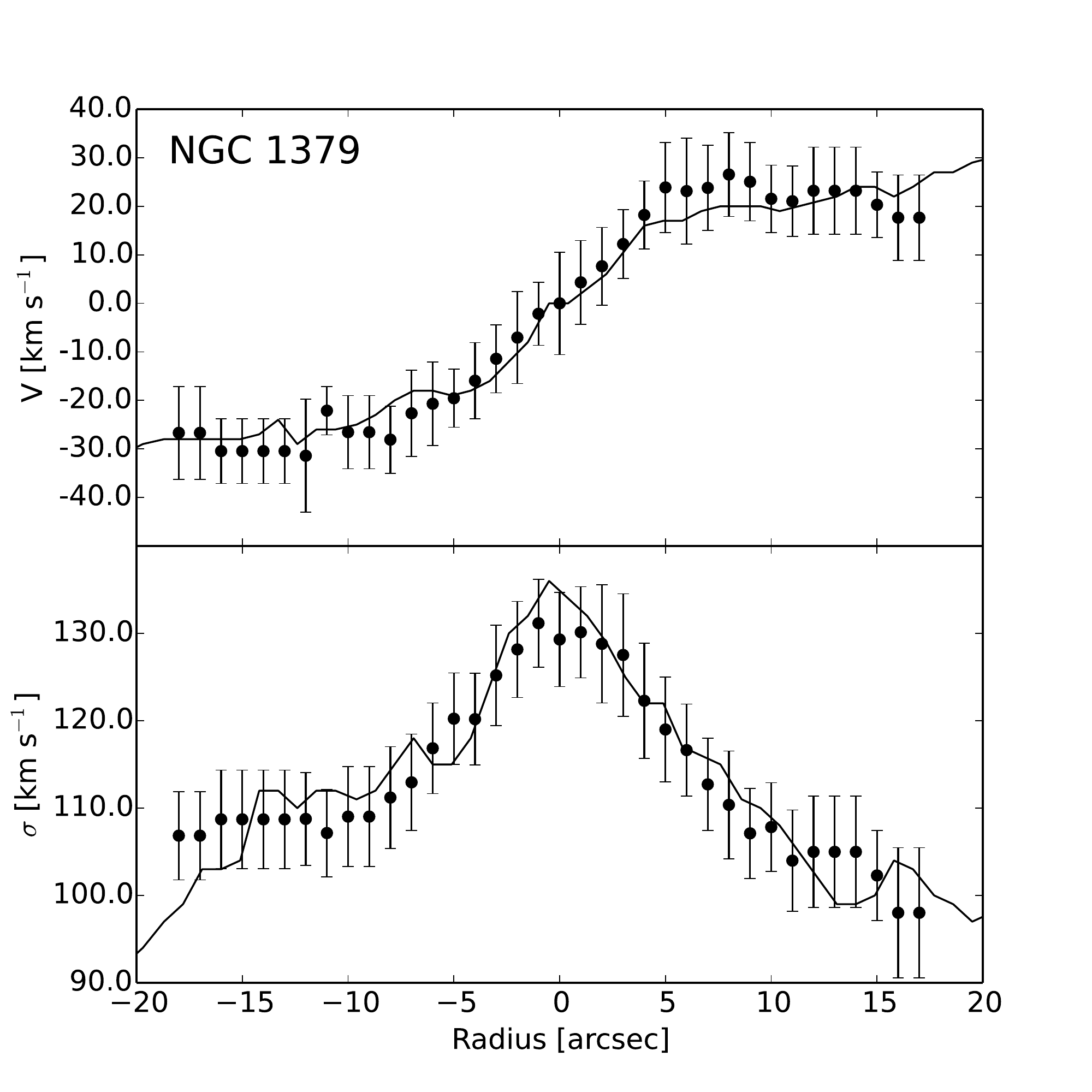}
\includegraphics[width=2.3in,clip,trim = 0 20 50 50]{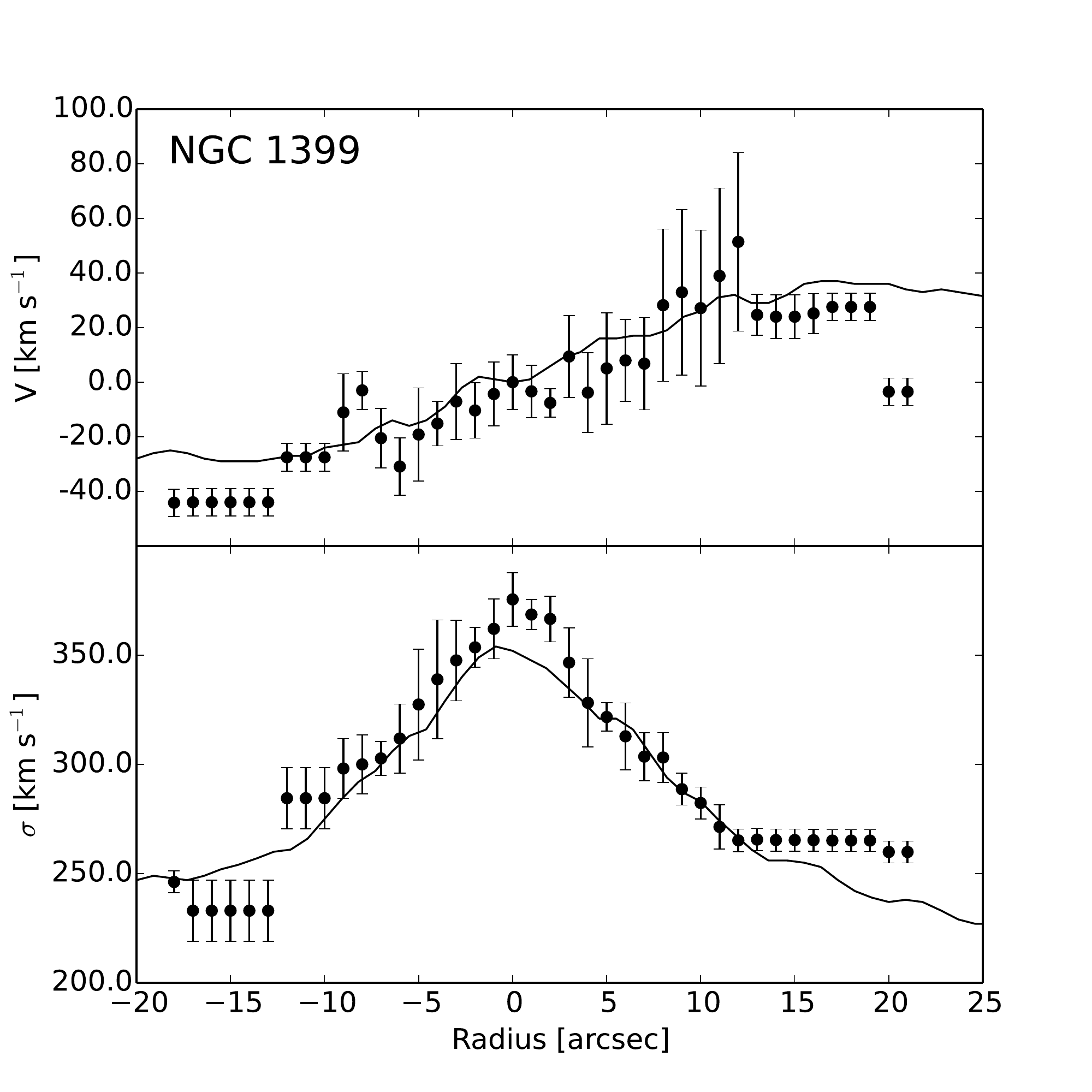}
\includegraphics[width=2.3in,clip,trim = 0 20 50 50]{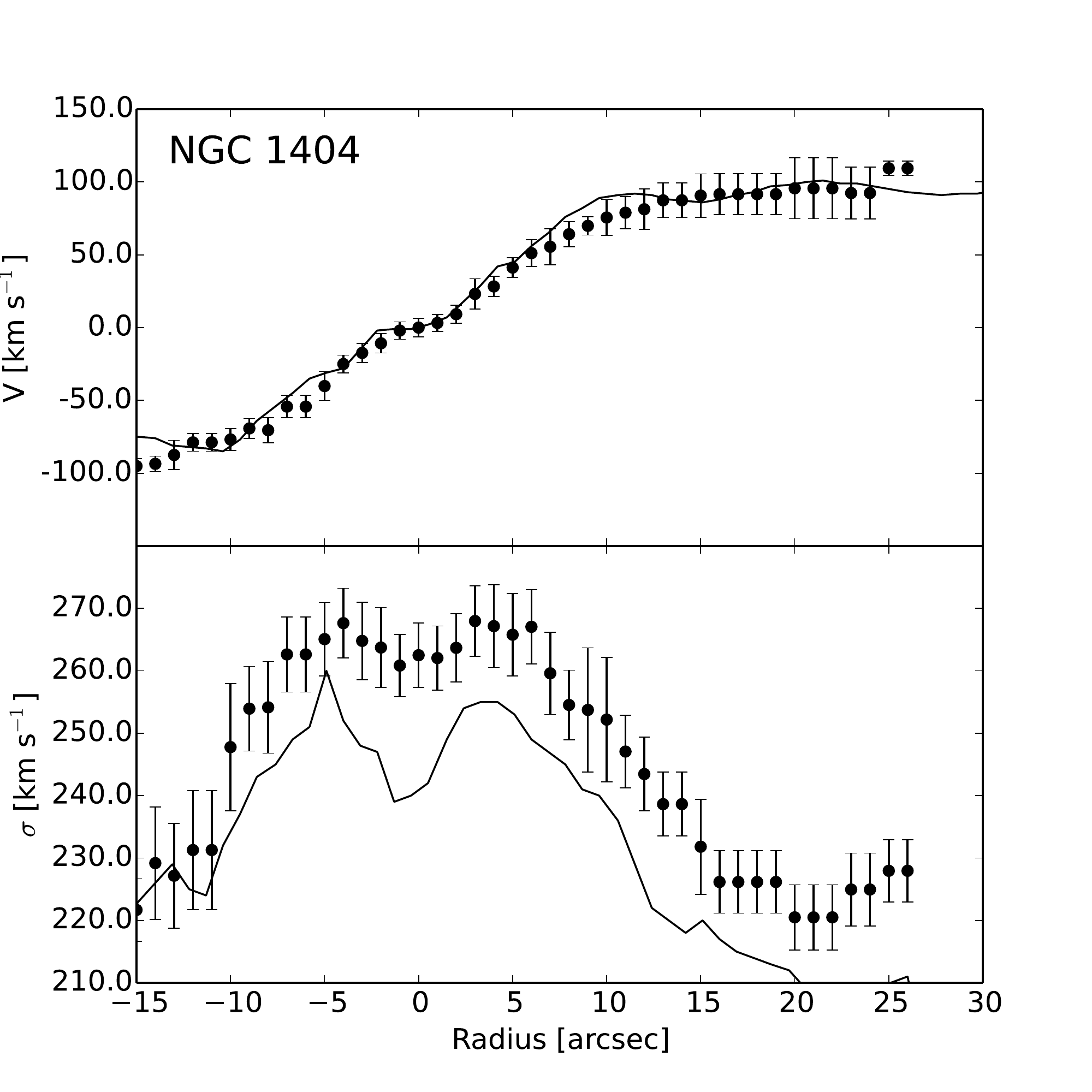}
\includegraphics[width=2.3in,clip,trim = 0 20 50 50]{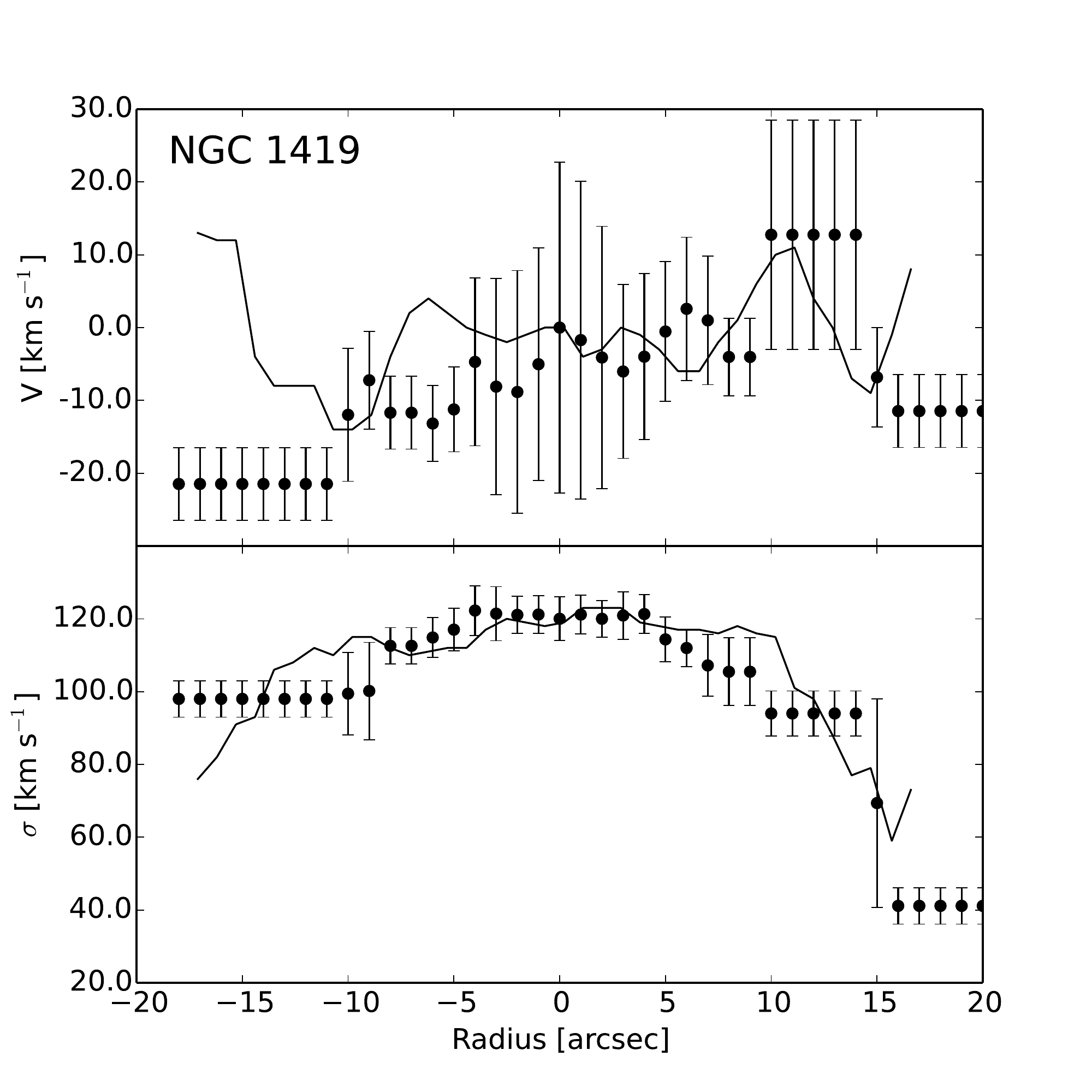}
\includegraphics[width=2.3in,clip,trim = 0 20 50 50]{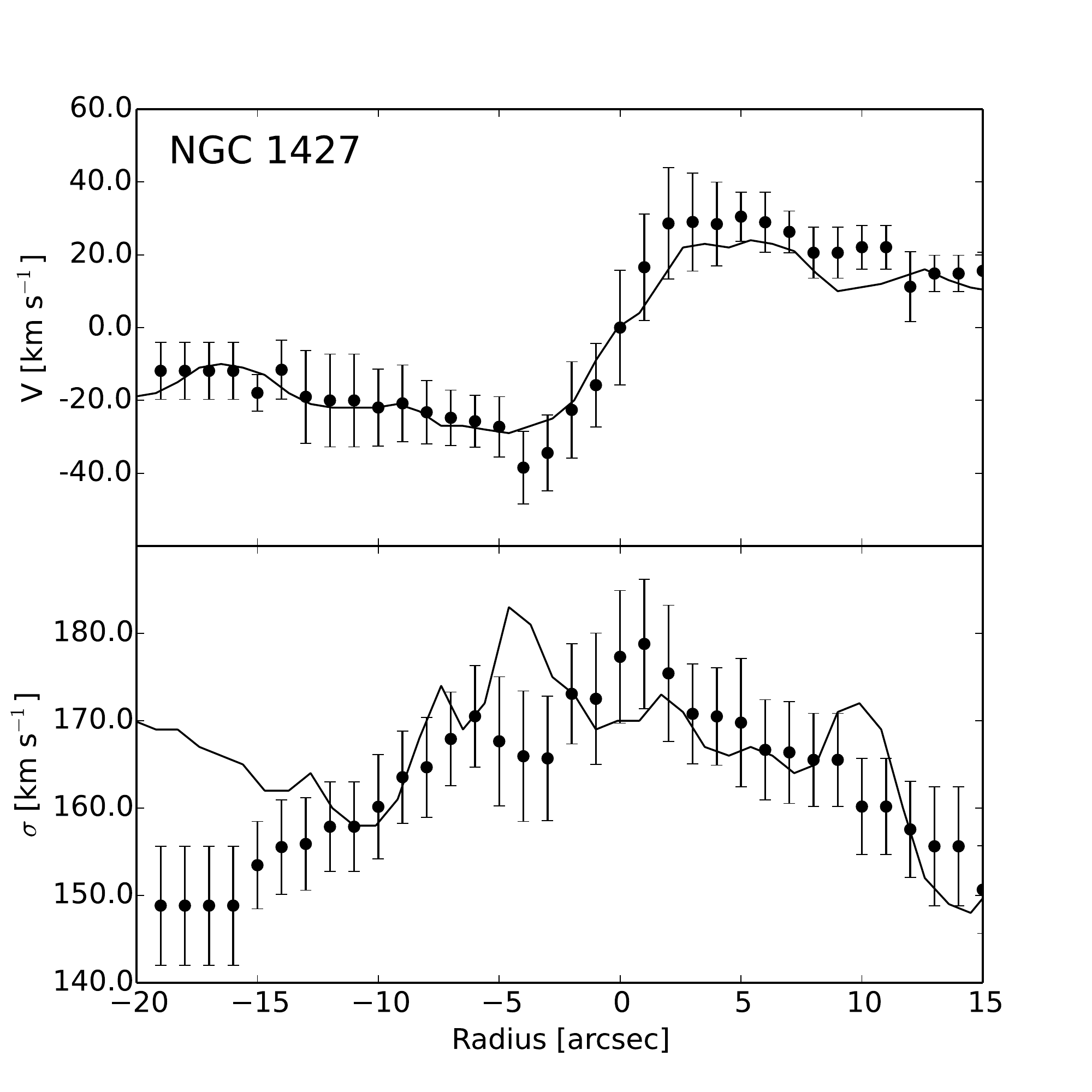}
\caption{Comparison between long-slit kinematic measurements of \citet[solid line]{Graham:1998} and those extracted from a matching mock long-slit aperture from our WiFeS kinematics for the 8 galaxies in common between the two samples.}
\label{fig:figurea2}
\end{figure*}

As discussed in Section \ref{sec:intro}, stellar kinematics derived from long-slit spectroscopic observations exist for many of the objects in our sample. Here we compare a subset of these literature measurements to kinematics derived from mock long slits applied to our WiFeS observations. In Figure \ref{fig:figurea2} we show the long-slit kinematic measurements of \citet[black lines]{Graham:1998} against our own kinematics (black points and error bars) extracted from an aperture matched to the \citet{Graham:1998} observations for the 8 galaxies in common between the two samples. In general the agreement is excellent, with the small variations likely attributable to differences in the atmospheric seeing between the two sets of observations, or to any residual mismatch with our mock long-slit aperture. The comparisons for NGC~1336 and NGC~1419 illustrate the value of IFS data for measuring galaxy kinematics. In these two cases the long-slit of \citet{Graham:1998} was aligned almost perpendicular to the direction of rotation, resulting in low measured rotation velocities. In contrast the WiFeS velocity maps presented in Figure \ref{fig:figurea1} clearly show ordered rotation in these two galaxies. Having noted that, \citet{Graham:1998} identified just 3 of the 12 brightest elliptical galaxies in the Fornax cluster are actually pressure supported systems (NGC~1399, NGC~1427 and FCC~335). This is in agreement with our findings, with the caveat that we did not observe as far down the luminosity function and therefore did not study FCC~335, the third pressure supported system. The relative absence of discs in these lower luminosity galaxies can however be seen from the $V/\sigma$ diagram of \citet[their figure~7]{DOnofrio:1995} and the bulge-disc decompositions of \citet{Graham:2003b}.

\label{lastpage}

\end{document}